\documentstyle[aps,twocolumn,eqsecnum]{revtex}

\input psfig.tex
\begin{document}
\draft
\wideabs{
\title{Quantum mechanical instabilities of Cauchy horizons in two
dimensions --- a modified form of the blueshift instability mechanism.}

\author{\'Eanna \'E.\ Flanagan}
\address{Cornell University, Newman Laboratory, Ithaca, NY
14853-5001.}
\maketitle

\begin{abstract}
There are several examples known of two dimensional spacetimes which
are linearly stable when perturbed by test scalar classical fields,
but which are unstable when perturbed by test scalar quantum fields.
We elucidate the mechanism behind such instabilities by considering minimally
coupled, massless, scalar, test quantum fields on general two
dimensional spacetimes with Cauchy horizons which are classically stable.
We identify a geometric
feature of such spacetimes which is a necessary condition for
obtaining a quantum mechanical divergence of the renormalized expected stress
tensor on the Cauchy horizon for regular initial states.  This feature
is the divergence of the affine parameter length of a certain one
parameter family of null
geodesics which lie parallel to the Cauchy horizon, where the affine
parameter normalization is determined by parallel transport along a
fixed, transverse null geodesic which intersects the Cauchy horizon.
(By contrast, the geometric feature of such spacetimes which underlies
classical blueshift instabilities is the divergence of a holonomy operator).
We show that the instability can be understood as a
``delayed blueshift'' instability, which arises from the infinite
blueshifting of an energy flux which is created locally and quantum
mechanically.
The instability mechanism applies both to chronology horizons
in spacetimes with closed timelike curves, and to the inner horizon in
black hole spacetimes like two dimensional Reissner-N\"ordstrom-de
Sitter.

\end{abstract}
}

\def\beq{\begin{equation}}
\def\endeq{\end{equation}}
\newtheorem{assumption}{Assumption}
\newtheorem{notion}{Cauchy horizon instability: meaning}

\narrowtext

\section{INTRODUCTION AND SUMMARY}
\label{fla-intro}


\subsection{Cauchy horizon instabilities : background and motivation}
\label{fla-overview}

The physical question addressed by the Haifa workshop can be
summarized as ``What is the generic physical nature of black hole
interiors?''.  Current attempts to answer this question involve
using all of the laws of physics which are well understood today,
including classical gravity and semiclassical gravity.  As remarked
by Valerie Frolov at the workshop, we theorists who study this issue are
fortunate in two separate ways:  First, we are protected from
possibly embarrassing confrontations with experimental data by the fact
that the gravitational singularities inside black holes are, very
likely, always hidden behind event horizons.  Second, some of the most
interesting and deep issues involving the singularities inside black
holes such as their possible traversability \cite{fla-Ori97} and such as the
information loss paradox \cite{fla-Preskill91} are ultimately obscured by
the ``Planck fog'' of Planck scale physics where both classical and
semiclassical gravity break down.   Rather than attempt to grapple
with these deep issues, we can justifiably throw up our hands once
Planck scale curvatures are reached and declare the subsequent
evolution to be beyond our purvue, given the absence of a well
understood theory of quantum gravity.

Nevertheless, it seems to me that it is still of interest to
investigate the structure of black hole interiors within the domains
of validity of classical and semiclassical gravity.  Black hole
interiors can in principle be probed experimentally, by observers who
venture inside the event horizon, and possibly even by external
observers if weak cosmic censorship turns out not to be valid.
Moreover, the structure of singularities even in the sub-Planckian
regime is a deep and complicated question, and is interesting both in
its own right and because understanding this structure is presumably a
necessary prerequisite to the eventual understanding of black holes in
full quantum gravity.

The study of black hole interiors is impeded by the fact that under
certain circumstances general relativity breaks down; indeed, it
predicts its own demise.  As is well known, given initial
data specified on some spacelike surface $\Sigma$, this breakdown can
take two forms.  First, in the maximal Cauchy evolution $D^+(\Sigma)$
of the initial data, the predicted gravitational field strength (curvature
scalars or components of curvature tensors on parallel propagated
bases along curves) can grow without limit.  This prediction is
presumably valid only up to the regime of Planckian curvatures, at
which point our understanding breaks down.  The prediction of
infinities or singularities is not particularly worrisome; similar
situations, where a continuum theory predicts its own breakdown, occur
in other contexts in physics.  A good example, as Amos Ori pointed out
during the workshop, is the formation of shocks in fluids.  There, the
hydrodynamic equations predict that the fluid variables diverge; in
actuality, the structure of shocks is determined by microscopic
physics for which the fluid approximation is invalid.  Presumably,
something similar occurs at gravitational singularities: their
structure is ruled by the as-yet-unknown laws of quantum gravity.

The second, well-known type of breakdown of general relativity is
where the maximal Cauchy evolution $D^+(\Sigma)$ of the initial data
is geodesically incomplete without any curvature singularities.
For example, if the initial data on $\Sigma$ were to consist of a
spherically symmetric, collapsing charged star, the maximal Cauchy
evolution would be a spacetime consisting of an interior solution
describing the star, together with an exterior solution consisting of
a portion of the Reissner-N\"ordstrom spacetime (the ``initial globally
hyperbolic region'').  In such spacetimes, certain observers, after a
finite amount of their proper time,  come to the ``edge'' of the
maximal Cauchy evolution; the theory fails to predict what such
observers subsequently measure.  Clearly this is a serious breakdown
of the theory \cite{fla-notecomplete}.

Mathematically, the breakdown is characterized by the fact that one
can extend the maximal Cauchy evolution, which we will
denote by $(M,g_{ab})$, to a larger spacetime $(M^\prime,
g_{ab}^\prime)$.  In any such larger spacetime, the future Cauchy
horizon $H^+(\Sigma)$ is the boundary of $M$ in $M^\prime$;
the breakdown is thus signified by the existence of a Cauchy horizon.
Note, however, that the extension spacetime is not uniquely determined
by the initial data and thus is not physical (even in situations where
a natural extension is determined by analytic continuation).

The disturbing aspect of this second type of breakdown is that it can
occur within the (apparent) domain of validity of general relativity,
entirely at low curvature scales, as in the Reissner-N\"ordstrom
example \cite{fla-notesecond}.  Of course, here we are assuming that
classical general relativity \cite{fla-noteterminology}
is a good approximation
at any point ${\cal P}$ in spacetime whenever the curvature is
sub-Planckian everywhere in the past lightcone of ${\cal P}$, which
seems like a reasonable assumption.
In order to highlight how disturbing such breakdowns
are, consider the hypothetical analogous situation in fluid hydrodynamics.
Suppose one were given a solution of the hydrodynamic equations in which
all of the lengthscales and timescales determined by the solution are
much larger than the relevant microscopic
lengthscales and timescales (so that the continuum approximation
should be good) but where the solution is nevertheless is
incomplete, cannot be uniquely extended, and fails to predict the
complete future evolution of the fluid.  Such a situation would be
paradoxical, and of course does not occur.

In the context of general relativity, if such breakdowns
were ubiquitous one would apparently be forced to abandon general
relativity
as a viable description of Nature even in the regime of low curvatures.
The traditional refuge of theorists has been to assert that such
breakdowns should be {\it non-generic}; that is, that only
``isolated'' initial data give rise to geodesically incomplete
spacetimes without curvature singularities.  In other words, Cauchy
horizons in spacetimes without singularities should always be
unstable. This is the essential content of the strong cosmic
censorship hypothesis \cite{fla-Wald84}.

The subject of this contribution to the proceedings is the
instability properties of Cauchy horizons in classical and
semi-classical gravity.  To summarize the above discussion, one of the
main motivations for studying the stability of Cauchy horizons is to
show that the breakdowns of general
relativity are not so serious as to render it not viable as a
description of Nature.  However, there are additional motivations.  In
attempting to determine the interior structure of black holes, one
finds that the well-known analytic solutions posses Cauchy horizons in
the interior, and in order to determine the generic interior structure
one must investigate solutions in a neighborhood of the solution with
the Cauchy horizon.  (Such investigations were
one of the focuses of the Haifa workshop.) Moreover, the stability of
Cauchy horizons is also relevant to the question ``Does Nature permit
the occurrence of closed timelike curves?'' \cite{fla-Thorne93}.

\subsection{Meaning of stability/instability}
\label{fla-defineunstable}

One would like to show that spacetimes with Cauchy horizons and
without curvature singularities are not generic, in the sense that
``generic'' perturbations to the initial data ${\cal I}$ for the
gravitational and matter fields on some initial Cauchy surface
$\Sigma$ will give rise to spacetimes without Cauchy horizons (i.e.,
inextendible spacetimes).  To
make the notion of genericity precise would involve defining a
topology and/or measure on the set of such initial data,
generic then meaning either ``all initial data sets ${\cal I}^\prime$
in some open set containing ${\cal I}$'' or ``all initial data sets
${\cal I}^\prime$ in some set whose complement is of measure zero''
\cite{fla-Wald97}.  It has become customary to say that a Cauchy horizon
is {\it unstable} when this non-genericity property is satisfied.

A somewhat different notion of instability is {\it linear instability}: a
Cauchy horizon is linearly unstable if generic perturbations to the
initial data (both matter and gravitational), when evolved forward
using the linearized Einstein-matter equations, yield a singularity of
the perturbed metric on the Cauchy horizon.  By continuity, one would
expect stability to imply linear stability, and thus linear
instability should be a sufficient condition for a true nonlinear
Cauchy horizon instability.  It need not be a necessary
condition as instabilities need not show up in linearized analyses,
but this has not occurred in most investigations to date in the context
of black holes.

In this contribution, we will use a still weaker notion of
instability.  We will call a Cauchy horizon {\it test field unstable}
if, when one evolves linearized test matter fields on the spacetime,
either classical or quantum mechanical, the stress-energy tensor of
the test field diverges on the Cauchy horizon (in the sense that
observers who cross the Cauchy horizon measure diverging stress tensor
components).   One would expect that test field instability should
imply linear instability, since the behavior of test matter fields
should presumably be similar to the behavior of linearized metric
perturbations, and also the test field's stress tensor should act as a
source for the leading order metric perturbation in a coupled
perturbation analysis.  Again, in cases that have been examined, test
field instability has been a good indicator of linear instability
\cite{fla-MellorMoss92}.  (For a detailed analysis of the relation between
various types of test field instability and nonlinear instability, see
Ref.~\cite{fla-Konkowski96} and references therein).

In the remainder of this contribution, we shall for the most part
adopt a conventional
abuse of terminology and abbreviate ``test field unstable'' as simply
``unstable''.  Thus, by stability or instability we shall {\it not}
mean the notion of full, nonlinear stability or instability discussed
above.  Our focus shall not be on understanding the effects of Cauchy
horizon instabilities, for which purpose one typically needs to
perform nonlinear analyses; rather we shall focus on trying to
understand when and why Cauchy horizon instabilities occur.
Also, we shall restrict attention to the simplified context of two
dimensional spacetimes.

\subsection{Purpose and overview of this contribution}
\label{fla-overview1}

Many investigations have found Cauchy horizons to be classically
linearly unstable, for example, Cauchy horizons in black holes in
asymptotically flat spacetimes
\cite{fla-SimpsonPenrose73,fla-ChandraHartle82}.  However, there is
some evidence for the existence of stable Cauchy horizons.
The first evidence for stable Cauchy horizons of which I am aware is
the work of Morris, Thorne and Yurtsever, who showed that the Cauchy horizons
in
wormhole spacetimes with closed timelike curves were classically test
field stable \cite{fla-MTY88}; their analysis was later generalized by
Hawking \cite{fla-Hawking92}.  However, the spacetimes in these analyses
were not solutions of Einstein's equation for a given matter model;
rather they were simply posited background spacetimes.
Also, the Cauchy horizon in the two dimensional version of the
Reissner-N\"ordstrom-deSitter spacetime is classically linearly test
field stable \cite{fla-BradyPoisson92}.

However, it has been shown in
Refs.~\cite{fla-DaviesMoss,fla-MarkovicPoisson95} that the two
dimensional Reissner-N\"ordstrom-deSitter spacetime is always
semiclassically unstable (see Sec.~\ref{fla-RNdeS} below for more
details).   Similarly, several researchers have shown that Cauchy
horizons in spacetimes with closed timelike curves are semiclassically
test field unstable
\cite{fla-KimThorne91,fla-Frolov91,fla-Others91,fla-KayWald97}, even
in those cases which are classically test field stable
\cite{fla-noteCTC}.

In the case of classically unstable Cauchy horizons, the physical
mechanism causing the instability is well understood --- it is just
the blueshifting of radiation to higher and higher energies
\cite{fla-SimpsonPenrose73,fla-ChandraHartle82,fla-BradyPoisson92}.
Moreover, one can identify a simple geometric feature of the
background spacetime --- whether or not the blueshift factor diverges
--- which allows one to predict whether or not the Cauchy horizon is
classically stable.

By contrast, our understanding of the nature of semiclassical
instabilities has been far less detailed.  We have had no simple and
general physical explanation of how or why such instabilities operate
in classically stable spacetimes.  The main purpose of this
contribution is to show that, in the simplified context of two
dimensional spacetimes, semiclassical instabilities can be understood
in a simple and intuitive way, and that one can identify a geometric
property of the background spacetime which allows one to predict
instabilities.

We examine conformally coupled, massless, scalar test quantum fields
on general two dimensional spacetimes with Cauchy horizons and without
singularities.  In such contexts, it is well known that the expected
stress tensor can be split in a unique way into the sum of two terms
[Eq.~(\ref{fla-generalsplit}) below]:  an ``initial data'' piece which
depends just on the initial values of the stress tensor on some
initial surface, and a ``locally generated'' piece which describes local
particle creation and/or vacuum polarization effects, and which
depends only on the spacetime geometry and not on the initial data.
Thus, there are two different types of semiclassical instabilities:
(i) Instabilities for which the ``initial data'' piece of the stress
tensor diverges on the Cauchy horizon.  This type of instability
arises classically as well as quantum mechanically; it is just the
blueshift instability.  (ii) Instabilities for which the ``locally
generated'' piece diverges on the Cauchy horizon.  We will call such
instabilities {\it locally created energy flux} instabilities.

We now define a particular geometric property of spacetimes which is
relevant to locally created energy flux instabilities.  In a
neighborhood of any point
${\cal P}$ on the Cauchy horizon, we
construct a family of null geodesics which lie parallel to the Cauchy
horizon.  We normalize the affine parameter $\lambda$ on these
geodesics by demanding that the vector field $d / d \lambda$ be
parallel transported along a fixed, transverse null geodesic $\Lambda$
which intersects the Cauchy horizon at ${\cal P}$.  Let $\Delta \lambda$
denote the total affine parameter length along any of these null geodesics
parallel to the Cauchy horizon, from $\Lambda$ back to the initial data
surface (see Fig.~\ref{fla-fig2} below).  Then, if $\Delta \lambda$
diverges as one moves closer and closer to the point ${\cal P}$ on the
Cauchy horizon, we will say that the spacetime has the property of
``divergence of affine parameter length''.

The main result of this contribution is that, under suitable mild
assumptions, the divergence of affine parameter length is a necessary
condition for a locally created energy flux instability.  Roughly speaking, if
the
total affine parameter length is bounded above and there are no
curvature singularities, then the locally small semiclassical
corrections do not have enough time to accumulate and cause a
divergence of the stress tensor.

The divergence of affine parameter length is not a
sufficient condition for semiclassical {\it test field} instabilities,
as there are locally flat spacetimes such as Misner space, which
satisfy the divergence of affine parameter length condition, but which
do not suffer from the locally created energy flux instability.  However,
locally flat
spacetimes are not generic; spacetimes close to Misner space but with
small amounts of curvature on the chronology horizon will suffer from
the locally created energy flux instability.  Hence, in suitable dynamical
two-dimensional versions of semiclassical gravity \cite{fla-note2D},
one would expect the locally created energy flux instability to be apparent in
second
order semiclassical perturbation theory about locally flat backgrounds
which satisfy the divergence of affine parameter length property.
We therefore conjecture that, quite generally, the
divergence of affine parameter length is a sufficient condition for
the full nonlinear instability of Cauchy horizons in suitable
dynamical two-dimensional versions of semiclassical gravity.

The properties of the blueshift and locally created energy flux
instability mechanisms are summarized and contrasted in Table
\ref{fla-table1} below.

The concept of divergence of affine parameter length meshes nicely
with our understanding of Cauchy horizon instabilities in the special
case of spacetimes with closed timelike curves
\cite{fla-KimThorne91,fla-Thorne93,fla-Hawking92,fla-KayWald97,fla-Klinkhammer92,fla-Visserbook,fla-Yurtsever91,fla-Sushkov97,fla-Krasnikov96,fla-Kay97}.
In such spacetimes, it is easy to see intuitively that the property of
divergence of affine parameter
length will always be satisfied: geodesics starting from points near the
Cauchy horizon (which will be a closed null geodesic in two
dimensions) will circle around very close to this closed null geodesic
many times before eventually making their way back to the initial data
surface.  We make this argument more precise in Sec.~\ref{fla-ctcs}
below \cite{fla-noteCTC1}.

Why should the property of divergence of affine parameter length be
relevant to divergences of the expected stress tensor?  There are two
different types of intuitive explanation which, although apparently
quite different, are not necessarily incompatible.
First, in spacetimes in which the curvature is low
everywhere, it is well known that classical general relativity coupled
to classical fields is locally a good approximation.  However, in a global
context,
small semiclassical corrections can ``accumulate'' and eventually
become important.  A good example of this is the Hawking evaporation
of macroscopic black holes, which takes place over long timescales.
Now when the property of divergence of affine parameter length is
satisfied, in some sense the quantum field perceives points very near
the Cauchy horizon to lie at ``asymptotically late times''.  Thus,
there is enough time for semiclassical corrections which are locally
small to accumulate and to become large at the Cauchy horizon.

The second intuitive explanation is that the locally created energy
flux instability can be understood as a modified type of blueshift
instability, a ``delayed blueshift instability''.  In the normal
blueshift instability mechanism, an energy
flux that is present in the initial data propagates from the initial
surface to a point near the Cauchy horizon, and in doing so suffers
some total net blueshift; this net blueshift becomes larger and larger
near the Cauchy horizon.  However, it is possible for the energy flux
to be first redshifted (frequency multiplied by some factor $F_{\rm
red} < 1$) and then blueshifted (frequency multiplied by some factor
$F_{\rm blue} >1$) in such a way that the total blueshift factor
$F_{\rm red} F_{\rm blue}$ is bounded above.  In such situations the
usual blueshift instability does not apply.  Nevertheless, if an
energy flux is {\it created} near the ``turning point'' where quanta
stop being redshift and start becoming blueshifted, then this locally
created energy flux suffers a net blueshift factor  $\sim F_{\rm blue}$
which is divergent, giving rise to a divergent stress tensor on the
Cauchy horizon.
In Sec.~\ref{fla-delayedblueshift} below we show that the above situation
always occurs when the property of divergence of affine parameter
length is satisfied in two dimensional spacetimes.  That is, the
divergence of the stress tensor is always caused by the infinite
blueshifting of a portion of the locally generated piece of the stress
tensor.

In Sec.~\ref{fla-fourdimensions} below we argue that
locally generated energy flux instabilities should also occur in four
dimensional spacetimes.  We also conjecture that, in the four
dimensional context, locally generated energy flux instabilities need
not always correspond to delayed blueshift instabilities.

\subsection{Organization of this contribution}

The organization of this contribution to the proceedings is as
follows.  Section
\ref{fla-foundations} is devoted to general analyses that underly
both the blueshift and locally created energy flux instabilities.  In
Sec.~\ref{fla-class} we define the class of spacetimes which we
analyze.
In Sec.~\ref{fla-nullbasissec} we define a basis of
null vectors which is adapted to the local geometry near a given point
on the Cauchy horizon.  This null basis serves as a convenient tool
throughout our calculations, and it allows us to avoid having to
introduce a coordinate system to describe the spacetime.  Using this
basis, we derive in Sec.~\ref{fla-STformula} a (well-known) general
formula for the expected stress tensor in terms of its initial values
on some initial surface $\Sigma$ [Eq.~(\ref{fla-basic1})].  That formula
exhibits the split of
the stress tensor, referred to above, into an ``initial data'' piece
plus a ``locally generated'' piece.  We show in
Sec.~\ref{fla-observersec} that all observers of bounded acceleration
measure finite energy densities while crossing the Cauchy
horizon if and only if the components of the expected stress tensor on
the null basis are finite.  Thus, in investigating the stability of
the Cauchy horizon, we can focus attention on the components of the
stress tensor on the null basis.

In Sec.~\ref{fla-blueshiftsec} we review and discuss the well-known
blueshift instability mechanism, in order to contrast it with the
locally created energy flux instability mechanism.  In
Sec.~\ref{fla-preliminary} we
give a very general definition of the blueshift factor, and recall that
it can be interpreted in terms of a holonomy operator around a certain
closed loop in spacetime, as illustrated in Fig.~\ref{fla-fig2}.  A
discussion of necessary and sufficient conditions for the instability
is given in Sec.~\ref{fla-gg}.  Finally, in
Sec.~\ref{fla-crossflowsec}, we review the well-known fact that the
instability acts only on radiation propagating parallel to the
Cauchy horizon and not to radiation which crosses the Cauchy horizon;
and we show that the radiation which crosses the Cauchy horizon is
always finite, semiclassically as well as classically.

Section \ref{fla-affinesec} is devoted to the locally created energy
flux instability
mechanism.  In Sec.~\ref{fla-necessarycondt} we show that the property
of divergence of affine parameter length is a necessary condition for
a locally created energy flux instability, when one assumes that the blueshift
factor
is globally bounded and one makes some other mild assumptions
about the spacetime and the initial slice $\Sigma$.  In
Sec.~\ref{fla-discussion} we explain that the divergence of affine
parameter length property is not a
sufficient condition for a linear instability (giving the
counterexample of Misner space), and conjecture that it should be a
sufficient condition for instabilities in a full, nonlinear analyses.
In Sec.~\ref{fla-fourdimensions}, we argue that some of the key ideas
which underly, in two dimensions, the classification of instabilities
into blueshift and locally created energy flux instabilities, should also
generalize to
four dimensions.  We speculate that the divergence of affine
parameter length {\it might} be relevant to Cauchy horizon
instabilities in four dimensions.

In Sec.~\ref{fla-delayedblueshift} we  show that the divergence of the
stress tensor is always caused by an infinite
blueshifting of a portion of the locally created piece of the stress
tensor, thus showing the instability mechanism can be understood as a
``delayed blueshift'' instability.

In Section \ref{fla-examplesec} we apply the general analyses of
Secs.~\ref{fla-foundations}, \ref{fla-blueshiftsec} and
\ref{fla-affinesec} to several different spacetimes and classes of
spacetimes, in order to clarify and illustrate the results.
In Sec.~\ref{fla-RNdeS}, we show that the property of divergence of
affine parameter length is satisfied in two-dimensional
Reissner-N\"ordstrom-deSitter
spacetimes, and reproduce the result of Markovi\'{c} and Poisson
\cite{fla-MarkovicPoisson95,fla-Poisson97} that such spacetime are
semiclassically unstable.  In Sec.~\ref{fla-ctcs} we show,
by adapting an argument due to Hawking, that two
dimensional spacetimes with closed timelike curves always
satisfy the property of divergence of affine parameter length.  A
particular example of such a spacetime, Misner space, is analyzed in
Sec.~\ref{fla-Misner}.  We
review the well-known fact that Misner space is blueshift unstable.
We also show that Misner space does {\it not} suffer from the
locally created energy flux instability mechanism; we argue, however, that
generic
spacetimes ``close to'' Misner space should suffer from the
instability.

Finally, section \ref{fla-conclusions} summarizes our main conclusions.
Appendix \ref{fla-generatorappendix} shows that the property of
divergence of affine parameter length will be satisfied at a point on
the Cauchy horizon whenever the generator of the Cauchy horizon
through that point has infinite affine parameter length in the past
direction.

We use units in which $G = c = \hbar = 1$, and we use the $(+,+,+)$
sign convention in the notation of Ref.~\cite{fla-MTW}.

\section{CAUCHY HORIZON STABILITY: FOUNDATIONAL ANALYSES}
\label{fla-foundations}

\subsection{Class of spacetimes and matter models}
\label{fla-class}

We start by specifying the class of spacetimes which we shall discuss.
We shall be interested in globally hyperbolic, two-dimensional
spacetimes $(M,g_{ab})$, i.e., spacetimes in which there exists a
partial Cauchy surface $\Sigma$ whose domain of dependence $D(\Sigma)$
is the entire spacetime $M$ \cite{fla-Wald84}.  The reason we restrict
attention to such spacetimes is the following.  Physically realistic
spacetimes should be obtainable by specifying initial data for the
gravitational and matter fields on some initial slice $\Sigma$ and by
solving the classical (or semiclassical) Einstein equation together
with the matter equations of motion in order to recover the entire
spacetime.  In this way one recovers the maximal Cauchy evolution of
the initial data, which will be just $D(\Sigma)$.  Thus, spacetimes
obtainable by this procedure will all be globally hyperbolic.

We will also assume that the spacetime $(M,g_{ab})$ is extendible to a
larger spacetime $(M^\prime,g_{ab}^\prime)$, and thus is geodesically
incomplete.  The spacetime $(M^\prime,g_{ab}^\prime)$ is useful from a
mathematical point of view --- using it one can define the future Cauchy
Horizon $H^+(\Sigma)$, which is just the boundary in $M^\prime$ of the
future domain of dependence $D^+(\Sigma)$.  Note, however, that the
extension is not unique and thus not physically meaningful.
Our arguments and conclusions will be independent of the choice of
extension $(M^\prime, g_{ab}^\prime)$.

The initial surface $\Sigma$ may either be spacelike, or it may
consist of two null segments as indicated in Fig.~\ref{fla-fig1} below.
Also, we do not restrict the topology of $\Sigma$: it may be compact
(topology of a circle) or non-compact (topology of the real line).  In
Sec.~\ref{fla-examplesec} below we discuss several specific examples of
spacetimes to which our analyses apply; most of these spacetimes will
be spatially non-compact, but one of them, Misner space, will be
spatially compact.

We will also assume that the spacetime $(M,g_{ab})$ does not have any
curvature singularities, either scalar or parallel propagated.  Thus,
we assume that (i) all local curvature invariants are
globally bounded on $(M,g_{ab})$, and (ii) there are no causal geodesic
curves along which the components of local curvature tensors on
parallel propagated bases diverge.  These assumptions are valid, for
example, for the initial globally hyperbolic regions $(M,g_{ab})$ of
the Reissner-N\"ordstrom and Reissner-N\"ordstrom-de Sitter spacetimes.

Our matter model will consist of a massless, minimally coupled
scalar quantum field ${\hat \Phi}$ on $(M,g_{ab})$.  We shall be concerned
with the expected stress-energy tensor $\langle T_{ab} \rangle$ of the
field ${\hat \Phi}$ in some quantum state in the vicinity of some point
${\cal P}$ on the Cauchy horizon.  For ease of notation, we will denote
this expected stress tensor simply as $T_{ab}$.

In this model, the test field stability of Cauchy horizons which
we are investigating should provide a good indication to the full, nonlinear
stability properties of two dimensional spacetimes, in the context of a
suitable two dimensional version of semiclassical gravity
\cite{fla-note2D}.   Alternatively, one can regard two dimensional,
semiclassical, test field instability results as implying that test
field instabilities are also likely in similar four dimensional,
spherically symmetric spacetimes \cite{fla-MarkovicPoisson95}; such four
dimensional test field instabilities would be directly relevant to
nonlinear instabilities in standard four dimensional semiclassical
gravity.

\subsection{General formula for stress tensor}
\label{fla-STformula}

The stress tensor $T_{ab}$ obeys the conservation equation
\beq
\nabla^a T_{ab} = 0
\label{fla-conservation}
\endeq
together with the trace anomaly equation \cite{fla-BirrellDavies82}
\beq
T_a^a = {1 \over 24 \pi} \, R,
\label{fla-traceanomaly}
\endeq
where $R$ is the two dimensional Ricci scalar.
As is well known \cite{fla-BirrellDavies82}, these conservation and trace
anomaly equations are sufficient to determine the evolution of
$T_{ab}$ from its initial value on $\Sigma$; one does not need to
specify the details of the quantum state, unlike the situation in four
dimensions.  We can therefore calculate the behavior of $T_{ab}$ near
the Cauchy horizon in terms of its initial data on $\Sigma$.

We split up the stress tensor into a traceless part and a trace part:
\beq
T_{ab} = {\hat T}_{ab} + {1 \over 48 \pi} \, R \, g_{ab},
\endeq
where $g^{ab} {\hat T}_{ab} = 0$.  It follows from
Eq.~(\ref{fla-conservation}) that the traceless part ${\hat T}_{ab}$ obeys
the equation
\beq
\nabla^a {\hat T}_{ab} = - {1 \over 48 \pi} \, \nabla_b R.
\label{fla-basic}
\endeq
This equation has a well posed initial value formulation; initial
values of ${\hat T}_{ab}$ determine uniquely its evolution.  The
general solution of Eq.~(\ref{fla-basic}) can be
written as
\beq
{\hat T}_{ab} = {\hat T}_{ab}^{({\rm initial\ data})} + {\hat
T}_{ab}^{({\rm locally\ generated})},
\label{fla-generalsplit}
\endeq
where ${\hat T}_{ab}^{({\rm initial\ data})}$ is the solution of the
homogeneous version of Eq.~(\ref{fla-basic}) with the same initial data on
$\Sigma$, and ${\hat T}_{ab}^{({\rm locally\ generated})}$ is the
solution of Eq.~(\ref{fla-basic}) with vanishing initial conditions.

We now derive explicit formulae for these two pieces of the general
solution.  These formulae are not new, but are usually derived and
expressed in a ``double-null'' coordinate systems $(u,v)$ in which the
metric is conformally flat:
\beq
ds^2 = - 2 e^{\sigma(u,v)} \, du dv;
\label{fla-conformalcoords}
\endeq
see, for example, Ref.~\cite{fla-BirrellDavies82}.  Here, however, we
derive and express the formulae using a basis of null vectors
$\{{\vec k}, {\vec l}\}$, which has the minor advantage that one does
not need to assume the global existence of a coordinate system of the
form (\ref{fla-conformalcoords}).

Let $\Gamma$ be any null geodesic in the spacetime, going from some point
${\cal Q}$ to some point ${\cal R}$.  Let $k^a$ denote the future
directed, null tangent to the geodesic, with associated affine
parameter $\lambda$, so that ${\vec k} = (d / d
\lambda)$.  Let $l^a$ be the vector field on $\Gamma$ which is null,
future directed and which satisfies $l^a k_a = -1$, from which it follows
that ${\vec l}$ is parallel transported along $\Gamma$.  Using the
relation
\beq
g_{ab} = -2 l_{(a} k_{b)}
\label{fla-gdecompos}
\endeq
in Eq.~(\ref{fla-basic}) yields
\beq
-2 k^{(a} l^{c)} \nabla_a {\hat T}_{cb} = - {1 \over 48 \pi} \nabla_b
R.
\endeq
Contracting this equation with $l^b$ yields
\beq
-k^a l^c b^b \nabla_a {\hat T}_{bc} - k^{(c} l^{b)} l^a \nabla_a {\hat T}_{cb}
 = - {1 \over 48 \pi} \, l^b \nabla_b R.
\label{fla-s1}
\endeq
The second term on the left hand side in Eq.~(\ref{fla-s1}) vanishes by
Eq.~(\ref{fla-gdecompos}).  Also the first term can be written as $-
d/d\lambda ({\hat T}_{bc} l^b l^c)$, since the vector ${\vec l}$ is
parallel transported along the geodesic.  Integrating with respect to
$\lambda$ and using ${\hat T}_{ab} l^a l^b = T_{ab} l^a l^b$ therefore yields
\beq
T_{ab} l^a l^b({\cal R}) = T_{ab} l^a l^b({\cal Q}) + {1 \over 48 \pi}
\int_{\cal Q}^{\cal R} d \lambda \,\,  l^a \nabla_a R.
\label{fla-basic1}
\endeq
Note that the first term in Eq.~(\ref{fla-basic1}) clearly corresponds to
the first term in Eq.~(\ref{fla-generalsplit}), and similarly for the
second term.

The formula (\ref{fla-basic1}) allows one to determine the entire stress
tensor from its initial value on $\Sigma$.  Namely, given any point
${\cal R}$ in $M$,
one can choose a pair of future directed null vectors ${\vec k}$ and
${\vec l}$ at
${\cal R}$ with $k^a l_a = -1$.  Then, the stress tensor at ${\cal R}$
can be written as
\beq
T_{ab}({\cal R}) = \rho l_a l_b + \sigma k_a k_b + {1 \over 24 \pi } \, R \,
g_{ab},
\label{fla-OnNullBasis}
\endeq
where $\rho = T_{ab} k^a k^b$ and $\sigma = T_{ab} l^a l^b$.  One can
determine $\sigma({\cal R})$ by shooting
off a past-directed null geodesic $\Gamma$ from ${\cal R}$ with
initial tangent $-{\vec k}$, extending it until it intersects the
initial surface $\Sigma$, parallel transporting ${\vec l}$ along
$\Gamma$, and then applying the formula
(\ref{fla-basic1}).  Similarly one can determine $\rho({\cal R})$ by
shooting off a null geodesic from ${\cal R}$ with initial tangent
$-{\vec l}$, and applying the formula (\ref{fla-basic1}) with ${\vec k}$
and ${\vec l}$ interchanged.

Note that, in the formula (\ref{fla-basic1}), one has the freedom to
perform the rescaling
\begin{eqnarray}
{\vec k} &\to& e^\mu {\vec k} \nonumber \\
\mbox{} {\vec l} &\to& e^{-\mu} {\vec l}
\label{fla-scalingfreedom}
\end{eqnarray}
where $\mu$ is a constant.  Both sides of
Eq.~(\ref{fla-basic1}) then get multiplied by $e^{- 2 \mu}$.  If one has a
family of geodesics (which corresponds to a specification of the basis
${\vec k}$, ${\vec l}$ in an open region), then $\mu$ can be any
function on the manifold $M$ satisfying $k^a \nabla_a \mu =0$.  In
terms of double null coordinates $(u,v)$
[cf.~Eq.~(\ref{fla-conformalcoords}) above], $\mu$ can be an arbitrary
function of the null coordinate $v$, when we adopt the convention
${\vec k} \propto \partial / \partial u$ and ${\vec l} \propto
\partial / \partial v$.

\subsection{A null basis adapted to the local geometry
near the Cauchy horizon}
\label{fla-nullbasissec}

Fix a point ${\cal P}$ in the Cauchy horizon $H^+(\Sigma)$.
We can construct a natural basis $\{{\vec k}$, ${\vec l}\}$ in the
intersection of the past light cone of ${\cal P}$ with a
neighborhood $U$ of ${\cal P}$, in which we resolve the
``gauge freedom'' (\ref{fla-scalingfreedom}) where $\mu = \mu(v)$, as
follows (see Fig.~\ref{fla-fig1}).  Let ${\vec k}({\cal P})$ be the
future-directed
tangent to the null generator of the Cauchy horizon at ${\cal P}$, with some
arbitrary choice of normalization.  Then there is a unique, null,
future directed vector ${\vec l}({\cal P})$ with
${\vec l} \cdot {\vec k} = -1$.  Let $\Lambda$ be the null geodesic
which starts from ${\cal P}$ with initial tangent $- {\vec l}$ and
extends into the past, and extend ${\vec l}$ and ${\vec
k}$ along
$\Lambda$ by parallel transport.  Finally, at an arbitrary point
${\cal R}$ on $\Lambda$ in $U$, shoot out a geodesic $\Gamma$
into the past with initial tangent $- {\vec k}$ and continue it until
it reaches that initial surface $\Sigma$, and extend ${\vec k}$ and
${\vec l}$ along $\Gamma$ by parallel transport \cite{fla-notebasis}.  The
resulting dreibein (set of basis vectors) $\{{\vec k}, {\vec l}\}$ is
unique up to transformations of the form (\ref{fla-scalingfreedom}) where
now $\mu$ is a constant.  Also, it follows from the construction that
Eq.~(\ref{fla-gdecompos}) holds in the domain of definition of
the null basis.

\subsection{Measurements made by observers crossing the Cauchy
horizon}
\label{fla-observersec}

Our aim is to characterize when the stress tensor $T_{ab}$ diverges at
the Cauchy horizon.  Now, as is well known, to detect divergences it
is insufficient in
general to examine coordinate invariant scalar quantities such as
$T_{ab} T^{ab}$ near the Cauchy horizon.  Instead, one must examine
the behavior of components of $T_{ab}$ with respect to parallel
propagated bases along curves of bounded proper
acceleration which cross the Cauchy horizon.  In other words, one must
examine what physical observers who cross the Cauchy horizon would
measure.  Now, we have constructed above a basis of null vectors which
is naturally adapted to the spacetime geometry in the vicinity of a
point ${\cal P}$ on the Cauchy horizon.  Therefore, one would expect
that the stress tensor is regular in the above parallel-propagated
sense at ${\cal P}$ if and only if the components of $T_{ab}$ on this
basis, i.e., the quantities $\sigma$ and $\rho$ defined by
Eq.~(\ref{fla-OnNullBasis}), are regular at ${\cal P}$.
In this section
we give a brief proof that this is indeed the case.  More precisely,
we shall show that the measured energy densities will be finite for
all observers of finite proper acceleration who cross the Cauchy
horizon at ${\cal P}$ if and only if both $\sigma$ and $\rho$ are bounded
above in a neighborhood of ${\cal P}$.

{\vskip 1cm}

{\psfig{file=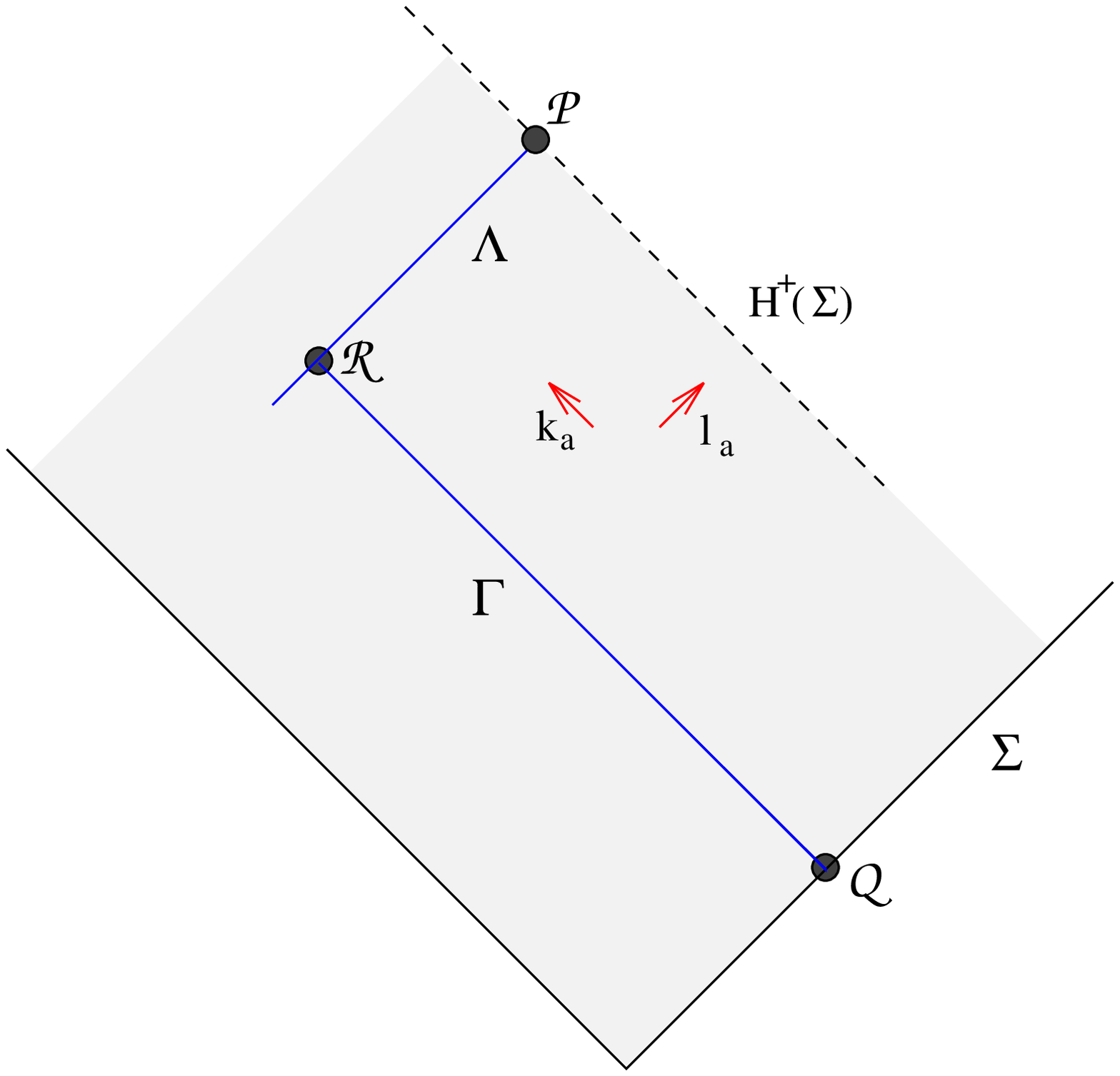,height=8cm,width=9cm,angle=-90}}
{\vskip 0.5cm}
\begin{figure}
\caption{An illustration of the construction outlined in the text.
The surface $\Sigma$ is the initial Cauchy surface, the shaded area
schematically represents the future domain of dependence
$D^+(\Sigma)$, and
$H^+(\Sigma)$ is the Cauchy horizon.  In the vicinity of any point
${\cal P}$ on the
Cauchy horizon, one chooses two future directed, null vectors ${\vec
k}$ and ${\vec l}$ with ${\vec k} \cdot {\vec l} = -1$ and with ${\vec k}$
normal to the Cauchy horizon, and one resolves the ambiguity in the
normalization of these vectors by starting at ${\cal P}$, parallel
transporting both vectors first along the geodesic $\Lambda$ and then
along the each geodesic $\Gamma$ starting from points ${\cal R}$ on
$\Lambda$.  The result is a dreibein $\{{\vec k},{\vec l}\}$ which is
adapted to the local geometry near ${\cal P}$ on the Cauchy horizon.
}
\label{fla-fig1}
\end{figure}

{\vskip 0.5cm}

Let ${\cal C}$ be the timelike worldline of an observer who passes
through ${\cal P}$.  Her velocity can be expressed as
\beq
{\vec u} = {1 \over \sqrt{2}} \left[ e^\chi {\vec l} + e^{-\chi} {\vec
k} \right]
\label{fla-hervelocity}
\endeq
for some $\chi = \chi(\tau)$, where $\tau$ is her proper time with
$\tau = 0$ at ${\cal P}$.  Now it follows from our construction of the
basis $\{{\vec k}, {\vec l}\}$ that
\begin{eqnarray}
\nabla_{\vec k} {\vec k} &=& \nabla_{\vec k} {\vec l} = 0 \nonumber \\
\mbox{} \nabla_{\vec l} {\vec l} &=& \kappa \, {\vec l} \nonumber \\
\mbox{}  \nabla_{\vec l} {\vec k} &=& - \kappa \, {\vec k}
\label{fla-2dimNP}
\end{eqnarray}
where $\kappa$ is a spin coefficient analogous to Newman-Penrose
quantity $\gamma$ in four dimensions.  Combining
Eqs.~(\ref{fla-hervelocity}) and (\ref{fla-2dimNP}) we obtain for the
acceleration
\beq
a^a = \left( {1 \over 2} e^{2 \chi} l^a -
k^a\right) \, \left( \kappa + \sqrt{2} e^{-\chi} u^b \nabla_b \chi
\right).
\endeq
Squaring this equation yields
\beq
{ d \chi \over d \tau} = \pm {1 \over \sqrt{2}} a(\tau) - {1 \over
\sqrt{2}} e^\chi \kappa,
\label{fla-boosteqn}
\endeq
where $a(\tau) \equiv \sqrt{a_a a^a}$.
It follows from Eq.~(\ref{fla-boosteqn}) that the boost parameter $\chi$
will be regular in a neighborhood of $\tau=0$, since by assumption
the geometry is regular in a neighborhood of ${\cal P}$, so $\kappa$
will be regular, and also the proper acceleration $a(\tau)$ is bounded
by assumption.

Next, the energy density $\rho_{\cal C} = T_{ab}\, u^a u^b$ that the observer
measures is given by, from Eqs.~(\ref{fla-OnNullBasis}) and
(\ref{fla-hervelocity}),
\beq
\rho_{\cal C} = {1 \over 2} \rho e^{-2 \chi} + {1 \over 2} \sigma e^{2
\chi} - {1 \over 24 \pi} R.
\endeq
The third term here is always bounded since the background geometry is
regular in a neighborhood of ${\cal P}$.
Now if both $\sigma$ and $\rho$ are finite at ${\cal P}$, then it
follows that $\rho_{\cal C}$ will be finite for all observers of
bounded acceleration, since for such observers $\chi$ is bounded.
Conversely, if either $\rho$ or $\sigma$ diverged at ${\cal P}$, then
there will be some choice of worldline ${\cal C}$ for which
$\rho_{\cal C}$ will be divergent \cite{fla-note1}.

To summarize, the Cauchy horizon will be stable if $\sigma$ and $\rho$
are finite on the Cauchy horizon, and unstable otherwise.

\subsection{Comparison between classical and semiclassical theories}
\label{fla-comparisonsec}

The classical version of the theory of a massless, conformally coupled
scalar field differs from the above semiclassical version in only two
respects:

\begin{itemize}

\item In the equation of motion (\ref{fla-basic}), the source term on the
right hand side is not present in the classical theory; it describes
local particle creation and/or vacuum polarization effects.
Correspondingly, in the split (\ref{fla-generalsplit}) of the stress
tensor into a ``locally generated'' piece and an ``initial data''
piece, the locally generated term is absent in the classical theory.

\item The set of allowed initial data for ${\hat T}_{ab}$ on $\Sigma$
is larger in the semiclassical theory than in the classical theory
\cite{fla-noteInitialData}.

\end{itemize}

Therefore, the piece ${\hat T}_{ab}^{({\rm initial\ data})}$ of the
general solution is {\it purely classical} --- it has exactly the same
behavior in semiclassical solutions as in classical solutions (except
for the greater freedom in initial data in semiclassical solutions).
In particular, if the Cauchy horizon is stable classically but
unstable quantum mechanically, the instability must be due to the
locally generated term ${\hat T}_{ab}^{({\rm locally\ generated})}$.
In the following two sections we turn to a discussion of the two different
instability mechanisms discussed in the Introduction: the well-known
blueshift instability which manifests itself as a
divergence of the term ${\hat T}_{ab}^{({\rm initial\ data})}$
and which can cause both classical and semiclassical instabilities,
and the locally created energy
flux instability mechanism which manifests itself as a divergence of the term
${\hat
T}_{ab}^{({\rm locally\ generated})}$ in Eq.~(\ref{fla-generalsplit}).

\section{THE BLUESHIFT INSTABILITY}
\label{fla-blueshiftsec}

The blueshift instability mechanism is responsible for the
classical instability of Cauchy horizons in charged and/or rotating
black hole spacetimes
\cite{fla-SimpsonPenrose73,fla-ChandraHartle82,fla-MellorMoss92,fla-ChambersMoss94},
which is nicely reviewed in the contributions by Poisson and Chambers
to this volume \cite{fla-Poisson97,fla-Chambers97}.  In this section,
we review this
mechanism in the language we have developed above, in order to
contrast the blueshift mechanism with the locally created energy flux
instability
mechanism we discuss in Sec.~\ref{fla-affinesec} below.
Our discussion will apply both to classical and quantum mechanical
analyses.

\subsection{Preliminary definitions and constructions}
\label{fla-preliminary}

As discussed in the Introduction, when we say that the Cauchy horizon
is unstable we mean that the stress tensor diverges on the Cauchy
horizon for generic, regular initial data on the initial Cauchy
surface $\Sigma$.  We now discuss what is a suitable meaning of
``regular'' in our context of very general two dimensional spacetimes.
To this end, we define a basis of null vectors $\{ {\vec k}_\Sigma,
{\vec l}_{\Sigma} \}$ on the initial surface $\Sigma$, according to
the following prescription: pick, at some point ${\cal Q}_0$ on
$\Sigma$, two future directed, null vectors ${\vec k}_\Sigma({\cal
Q}_0)$ and ${\vec l}_{\Sigma}({\cal Q}_0)$ with
${\vec k}_\Sigma({\cal Q}_0) \cdot {\vec l}_{\Sigma}({\cal Q}_0) = -1$,
and extend this basis to all of $\Sigma$ by parallel transport.
If $\Sigma$ has the topology of a circle, as is the case
for Misner space discussed in Sec.~\ref{fla-Misner} below, this definition
of the basis $\{ {\vec k}_\Sigma, {\vec l}_\Sigma \}$ is ambiguous.
We resolve the ambiguity by demanding that the
the basis be continuous at all points on $\Sigma$ except possibly at
${\cal Q}_0$; a non-unit holonomy around $\Sigma$ will yield a
discontinuity at ${\cal Q}_0$.  The resulting basis
$\{ {\vec k}_\Sigma, {\vec l}_\Sigma \}$ is unique up to an overall
constant boost of the form (\ref{fla-scalingfreedom}).

We will say that initial data for $T_{ab}$ on $\Sigma$ is {\it
regular} if the components $T_{ab} l^a_\Sigma l^b_\Sigma$ and $T_{ab}
k^a_\Sigma k^b_\Sigma$ of the stress tensor are continuous (at all
points except possibly ${\cal Q}_0$) and bounded.
It is clear that all physically reasonable initial data should be
regular in this sense.  In some spacetimes, the set of ``physically
reasonable'' initial data will be a proper subset of the set of regular
initial data (due, eg, to asymptotic fall-off conditions); see the
discussion of various examples of spacetimes in Sec.~\ref{fla-examplesec}.

We now extend the basis $\{ {\vec k}_\Sigma, {\vec l}_{\Sigma} \}$ to
the entire spacetime $(M,g_{ab})$ by parallel transporting along null
geodesics parallel to ${\vec k}_\Sigma$.   This basis will be related
to the basis $\{ {\vec k}, {\vec l} \}$ discussed in
Sec.~\ref{fla-nullbasissec} above by a spacetime dependent boost:
\begin{eqnarray}
{\vec k} &=& e^{-\Psi} {\vec k}_\Sigma \nonumber \\
{\vec l} &=& e^\Psi {\vec l}_\Sigma.
\label{fla-Psidef}
\end{eqnarray}
Since the vector fields ${\vec k}$ and ${\vec k}_\Sigma$ are geodesic,
we have $k^a \nabla_a \Psi =0$, {\it i.e.}, $\Psi$ depends only on one
of the two null coordinates $(u,v)$.  For convenience, we will always
choose the boost-normalization of the basis $\{{\vec k}, {\vec l} \}$
so that $\Psi({\cal Q}_0) = 0$.

{\vskip 1cm}

{\psfig{file=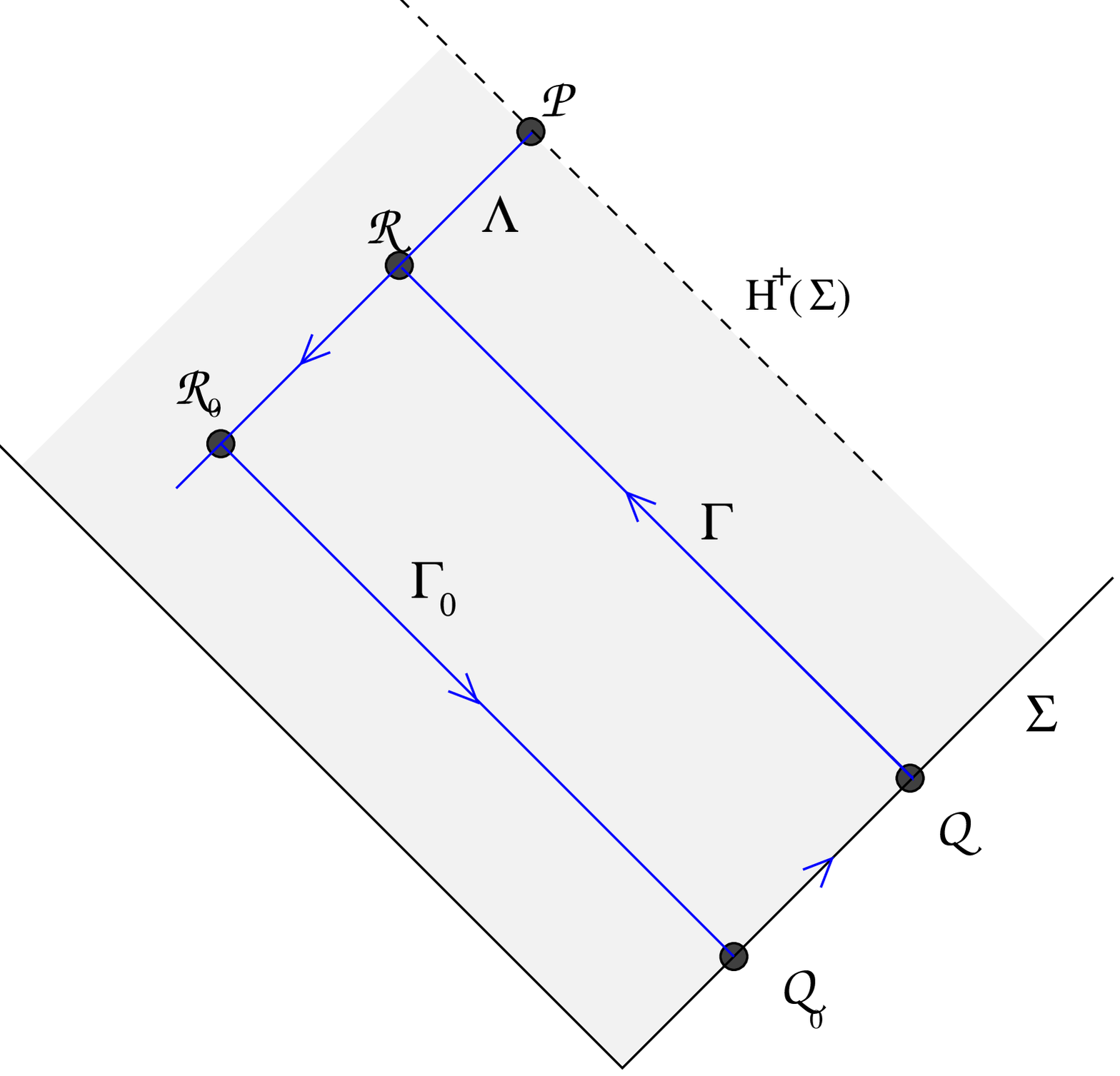,height=8cm,width=8cm,angle=-90}}
\begin{figure}
\caption{
A specialization of the construction shown in Fig.~\ref{fla-fig1} to
illustrate the difference between the blueshift and locally created
energy flux instability
mechanisms.  Here $\Sigma$ is the initial Cauchy surface, ${\cal P}$
is a point on the future Cauchy horizon $H^+(\Sigma)$, and ${\cal R}$
and ${\cal R}_0$ are points on the null geodesic $\Lambda$ emanating
from the Cauchy horizon at ${\cal P}$, which are connected to points
${\cal Q}$ and ${\cal Q}_0$ on $\Sigma$ via null geodesics.  The {\it
blueshift instability} mechanism is characterized by a divergence of
the total holonomy around the loop ${\cal R} {\cal R}_0 {\cal Q}_0
{\cal Q}$ as ${\cal R} \to {\cal P}$.  By contrast, what we call the
{\it locally created energy flux instability} mechanism is characterized by the
divergence of the affine
parameter length of the geodesic ${\cal R} {\cal Q}$ as ${\cal R} \to
{\cal P}$, where the normalization of the affine parameter $\lambda$
is such that $d/d \lambda$ is parallel transported along the geodesic
$\Lambda$.  This property of divergence of affine parameter length is a
necessary but not sufficient condition for a locally created energy
flux instability.
}
\label{fla-fig2}
\end{figure}

{\vskip 0.5cm}

Note that the quantity $\Psi$ can be understood as characterizing a
holonomy operator around a closed loop in spacetime, in the special
case where the topology of $\Sigma$ is that of the real line (which
excludes examples like Misner space).  More specifically, consider the
construction of the basis $\{ {\vec k}, {\vec l} \}$ described in
Sec.~\ref{fla-nullbasissec} above.  That construction describes a family
of null geodesics $\Gamma$ joining points ${\cal Q}$ on the initial
slice $\Sigma$ to
points ${\cal R}$ on the geodesic $\Lambda$ which emanates from the
Cauchy horizon at the point ${\cal P}$ (see Fig.~\ref{fla-fig2}).
Pick a particular null geodesic $\Gamma_0$ joining some points ${\cal
R}_0$ and ${\cal Q}_0$.  Then it follows that at any point ${\cal R}$
on $\Lambda$, the holonomy around the loop ${\cal R} \, {\cal R}_0 \,
{\cal Q}_0 \, {\cal Q} \, {\cal R}$ indicated in Fig.~\ref{fla-fig2} is
simply the following boost with rapidity parameter $\Psi({\cal R})$:
\beq
\alpha \, {\vec k}({\cal R}) + \beta \, {\vec l}({\cal R}) \to
\alpha \, e^{\Psi({\cal R})} \, {\vec k}({\cal R}) + \beta \, e^{-\Psi({\cal
R})} \,  {\vec l}({\cal R}).
\label{fla-holonomy}
\endeq
Here $\alpha$ and $\beta$ are arbitrary real numbers.  To derive
Eq.~(\ref{fla-holonomy}), consider starting at ${\cal R}$ with the
vector ${\vec
k}({\cal R})$.  When one parallel transports this vector to ${\cal
R}_0$ along $\Lambda$ and then to ${\cal Q}_0$ along $\Gamma_0$, the
result is just ${\vec k}({\cal Q}_0)$ by the definition of the
basis $\{ {\vec k}, {\vec l} \}$, which equals ${\vec
k}_\Sigma({\cal Q}_0)$ from Eq.~(\ref{fla-Psidef}) and
the fact that $\Psi({\cal Q}_0)=0$.  When one then parallel transports
this vector along $\Sigma$ to ${\cal Q}$ and then along $\Gamma$ back
to ${\cal R}$, the result is ${\vec k}_\Sigma({\cal R})$ by the
definition of the basis $\{ {\vec k}_\Sigma, {\vec l}_\Sigma \}$,
which by Eq.~(\ref{fla-Psidef}) is the same as $e^{\Psi({\cal R})} \,
{\vec k}({\cal R})$.  A similar argument applies when one starts at
${\cal R}$ with ${\vec l}({\cal R})$, and one obtains in this way
Eq.~(\ref{fla-holonomy}).

Note also that the blueshift factor $e^\Psi$ can be understood as the
ratio
of differential proper times of freely falling observers near ${\cal
P}$ and on $\Sigma$, as explained in detail in
Refs.~\cite{fla-BradyPoisson92,fla-Chambers97}.

\subsection{Instability mechanism}
\label{fla-gg}

We now describe the blueshift instability mechanism.  Spacetimes in
which this instability operates have the property that, for some point
${\cal P}$ on the Cauchy horizon, the ``blueshift factor''
$e^{\Psi({\cal R})}$ diverges as the point ${\cal R}$ approaches
${\cal P}$.  From Eqs.~(\ref{fla-OnNullBasis}), (\ref{fla-basic1}) and
(\ref{fla-Psidef}) we find
that the quantity $\sigma$ (describing radiation propagating parallel
to the Cauchy horizon) evaluated at the point ${\cal R}$ near the
Cauchy horizon is given by
\begin{eqnarray}
\sigma({\cal R}) &=& e^{2 \Psi({\cal R})} \, \sigma_\Sigma\left[{\cal
Q}({\cal R})\right] \nonumber\\
\mbox{} && + (\mbox{locally\ generated\ term}).
\label{fla-blueshiftans}
\end{eqnarray}
Here, ${\cal Q}({\cal R})$ denotes the unique point ${\cal Q}$ on
$\Sigma$ determined by ${\cal R}$ according to the construction
described in Sec.~\ref{fla-nullbasissec}, and $\sigma_\Sigma \equiv T_{ab}
l^a_\Sigma l^b_\Sigma$ is the initial data on $\Sigma$ which
we have assumed is bounded.  In this section we will
ignore the locally generated term in Eq.~(\ref{fla-blueshiftans}); it will
be absent in a classical treatment, and even in a semiclassical
treatment, a divergence of the first term in Eq.~(\ref{fla-blueshiftans})
should be sufficient to produce an instability \cite{fla-notquite}.

{}From Eq.~(\ref{fla-blueshiftans}) the quantity $\sigma({\cal R})$ will
diverge at ${\cal P}$ unless the initial data $\sigma_\Sigma[{\cal
Q}({\cal R})]$ goes to zero as ${\cal R} \to {\cal P}$ faster than the
divergence of the blueshift factor $e^{\Psi({\cal R})}$.  Thus,
whether or not the spacetime is unstable depends on the precise
specification of the class of ``physically reasonable'' initial data
on $\Sigma$ (cf.~Sec.~\ref{fla-preliminary} above).

There are several possibilities for the behavior of ${\cal Q}({\cal R})$
on $\Sigma$ as ${\cal R} \to {\cal P}$:

\begin{itemize}

\item When $\Sigma$ is non-compact, it can happen that ${\cal Q}({\cal
R}) \to \infty$ as ${\cal R} \to {\cal P}$.  Then the behavior of
$\sigma_\Sigma[{\cal Q}({\cal
R})]$ depends on the asymptotic fall-off behavior of the initial data.
In the case of classically unstable black hole spacetimes, the fall
off rate of $\sigma_\Sigma[{\cal Q}({\cal R})]$ for physically
reasonable initial
data is sufficiently slow that, although $\sigma_\Sigma[{\cal Q}({\cal
R})] \to 0$, the quantity (\ref{fla-blueshiftans}) still diverges
\cite{fla-BradyPoisson92}.

\item It can happen that ${\cal Q}({\cal R}) \to {\cal Q}_1$ as ${\cal
R} \to {\cal P}$, where ${\cal Q}_1$ is some fixed point on $\Sigma$
\cite{fla-example2}.  In this case, generic initial data will have
$\sigma_\Sigma({\cal Q}_1) \ne 0$, so the spacetime will be unstable.

\item In the case where $\Sigma$ is compact, it can happen that ${\cal
Q}({\cal R})$ ``circles around and around'' $\Sigma$ as ${\cal R} \to
{\cal P}$, as occurs for example in Misner space.  In this case also,
it follows from Eq.~(\ref{fla-blueshiftans}) that $\sigma({\cal R})$ will
be unbounded near ${\cal P}$ for all non-zero initial data, and thus
the spacetime will be unstable.

\end{itemize}

To summarize, when the background spacetime $(M,g_{ab})$ has the
property that the holonomy-like quantity $e^{\Psi({\cal R})}$ diverges,
and when in the non-compact case the fall-off rate of the class of
``physically reasonable'' initial data is sufficiently slow,
then the spacetime will be (classically and quantum mechanically)
unstable.  The condition on the fall-off rate of the initial data
depends on the spacetime and on the choice of initial surface
$\Sigma$; it is satisfied in black hole spacetimes \cite{fla-caveat1}.

\subsection{Radiation crossing the Cauchy horizon}
\label{fla-crossflowsec}

So far in this section we have considered only the term $\sigma k_a
k_b$ in Eq.~(\ref{fla-OnNullBasis}), which describes radiation propagating
parallel to the Cauchy horizon.  What of the other term $\rho l_a
l_b$ in that equation?  This term describes radiation which crosses
the Cauchy horizon.  We now show that such radiation can never give
rise to an instability by showing that the quantity $\rho$ is always
bounded near the point ${\cal P}$.

Extend the geodesic $\Lambda$ into the past until it intersects the
initial surface $\Sigma$ at some point ${\cal S}$.  Then, since the
vectors ${\vec k}$ and ${\vec l}$ are both parallel transported along
$\Lambda$, it follows from Eq.~(\ref{fla-basic1}) with ${\vec k}$ and
${\vec l}$ interchanged that
\beq
\rho({\cal R}) = (T_{ab} k^a k^b)({\cal S}) + {1 \over 48 \pi}
\int_{\cal S}^{\cal R} d \xi \,\,  k^a \nabla_a R,
\label{fla-crossflow0}
\endeq
where $\xi$ is an affine parameter along $\Lambda$ such that ${\vec l}
= d / d\xi$.  The first term in Eq.~(\ref{fla-crossflow0}) is just
\beq
e^{-2 \Psi({\cal S})} \, \rho_\Sigma({\cal S}),
\label{fla-crossflow1}
\endeq
where $\rho_\Sigma \equiv T_{ab} k^a_\Sigma k^b_\Sigma$, from
Eq.~(\ref{fla-Psidef}).  Now the point ${\cal S}$ does
not vary as ${\cal R} \to {\cal P}$, and thus the quantity
(\ref{fla-crossflow1}) is fixed and finite as ${\cal R} \to {\cal P}$.
The second term in Eq.~(\ref{fla-crossflow0}) will converge to the finite
quantity
\beq
{1 \over 48 \pi}
\int_{\cal S}^{\cal P} d \xi \,\,  k^a \nabla_a R
\endeq
as ${\cal R} \to {\cal P}$, and so $\rho({\cal R})$ is bounded as
${\cal R} \to {\cal P}$.

Thus, when the blueshift factor is finite, it follows from
Eq.~(\ref{fla-blueshiftans}) and the analysis of this subsection that the
spacetime is stable except possibly for a divergence of the
semiclassical, locally generated piece of $\sigma({\cal R})$.  We now turn
to a discussion of this possibility.

\section{THE LOCALLY CREATED ENERGY FLUX INSTABILITY}
\label{fla-affinesec}

In this section we discuss a second instability mechanism which
can cause a divergence of the second term ${\hat T}_{ab}^{({\rm
locally\ generated})}$ in Eq.~(\ref{fla-generalsplit}).  This mechanism
is a purely quantum mechanical effect as the term ${\hat T}_{ab}^{({\rm
locally\ generated})}$ is absent in classical analyses.

Let us focus attention on spacetimes in which the blueshift factor
$e^\Psi$ is globally bounded, and in which therefore the blueshift
instability mechanism does not operate.  For example, the two dimensional
Reissner-N\"ordstrom-de Sitter spacetime has this property in a
certain region of parameter space \cite{fla-BradyPoisson92}, when the
initial surface $\Sigma$ is chosen to lie outside the event horizon (see
Sec.~\ref{fla-RNdeS} below).  Such spacetimes will be classically stable
but may be semiclassically unstable.

{}From Eq.~(\ref{fla-basic1}), the locally generated piece of $\sigma$ is
given by
\beq
\FL
\sigma^{({\rm locally\ generated})}({\cal R}) = {1 \over 48 \pi}
\int_{{\cal Q}({\cal R})}^{\cal R} d \lambda
\,\,  l^a \nabla_a R.
\label{fla-figureofmerit}
\endeq
Now by assumption the background geometry is nonsingular, so the Ricci
scalar $R$ is bounded; however the quantity $l^a \nabla_a R$ may still
be unbounded and may give rise to a divergence of the integral
(\ref{fla-figureofmerit}) as ${\cal R} \to {\cal P}$.
Alternatively, one might imagine that
the integrand $l^a \nabla_a R$ could be bounded, but that
the total affine parameter length in $\lambda$ between ${\cal Q}({\cal
R})$ and ${\cal R}$ could diverge as ${\cal R} \to {\cal P}$ giving rise
to a divergence of the integral.  In Sec.~\ref{fla-examplesec} below, we
examine several spacetimes for which the integral
(\ref{fla-figureofmerit}) diverges, and we find that in these example
spacetimes, {\it both} the integrand and the total affine parameter
length diverge.

We will say that a spacetime has the property of ``divergence of
affine parameter length'' if, for some point ${\cal P}$ on the Cauchy
horizon,
\beq
\Delta \lambda ({\cal R}) \to \infty \ \ \ \ \ {\rm as} \ \ \ \  {\cal
R} \to {\cal P}.
\label{fla-DAPL}
\endeq
Here $\Delta \lambda({\cal R})$ is the affine parameter length of the
null geodesic from ${\cal R}$ to ${\cal Q}({\cal R})$, where the
affine parameter $\lambda$ is normalized according to ${\vec k} = d /
d\lambda$ (see Fig.~\ref{fla-fig2} above).  In Appendix
\ref{fla-generatorappendix} we show that the property
(\ref{fla-DAPL}) will hold if the generator of the Cauchy horizon through
the point ${\cal P}$ has infinite affine parameter length towards the
past.  [Note that, although any such generator must be inextendible in
the past direction \cite{fla-Wald84}, it may have finite affine parameter
length in the past direction if the spacetime
$(M^\prime,g_{ab}^\prime)$ is geodesically incomplete; see, for
example, Fig.~8.2 of Ref.~\cite{fla-Wald84}.]

\subsection{A necessary condition for instability}
\label{fla-necessarycondt}

In this subsection section we prove that when one makes
certain mild assumptions about the spacetime $(M,g_{ab})$ and the
initial slice $\Sigma$,  the property of
divergence of affine parameter length is a {\it necessary} condition
for an instability of the Cauchy horizon mediated by a divergence of
the quantity (\ref{fla-figureofmerit}).

A precise statement of our result is the following.  Assume that
(i) The spacetime $(M,g_{ab})$ is non-singular in the sense
discussed in Sec.~\ref{fla-class} above;  (ii) The holonomy-like quantity
$\Psi({\cal R})$ is globally bounded;  (iii) The total affine
parameter length $\Delta \lambda({\cal R})$ is globally bounded by
some maximum $\Delta \lambda_{\rm max}$; (iv) The initial slice
$\Sigma$ is regular in the sense that the quantity $l^a_\Sigma
\nabla_a R$ is bounded on $\Sigma$.  Then the quantity
(\ref{fla-figureofmerit}) is bounded and thus the Cauchy horizon is
stable.  In other words, assuming the conditions (i), (ii) and (iv),
the divergence of affine parameter length is a necessary condition for
an instability.

Assumption (iv) is only relevant when $\Sigma$ is
non-compact; it is satisfied automatically in compact cases like
Misner space.  In the non-compact case, the condition (iv) is a
reasonable assumption
as the vector ${\vec l}_\Sigma$ is parallel transported along
$\Sigma$, and moreover if $\Sigma$ is asymptotically spacelike and the
spacetime is asymptotically flat, then $R$ will go to zero at large
distances.  The condition should thus be satisfied by slices
$\Sigma$ which are asymptotically null or asymptotically spacelike in
spacetimes that are asymptotically flat.

We now turn to a proof of the above result.  Consider the quantity
$k^a \nabla_a \, l^b \nabla_b R$, which we can write as
\begin{eqnarray}
k^a \nabla_a \, l^b \nabla_b R &=& k^a l^b \nabla_a \nabla_b R + ( k^a
\nabla_a l^b) \nabla_b R \nonumber \\
\mbox{} &=& - {1 \over 2} \Box R.
\end{eqnarray}
Here the second term on the first line vanishes by Eq.~(\ref{fla-2dimNP}),
and the second equality follows from Eq.~(\ref{fla-gdecompos}).  Now
integrating with respect to $\lambda$ along the geodesic $\Gamma$ in
Fig.~\ref{fla-fig1}, using $k^a \nabla_a = d/d\lambda$ and
Eq.~(\ref{fla-Psidef}),
we obtain
\begin{eqnarray}
(l^a \nabla_a R)({\cal R}^\prime) &=& (l^a \nabla_a R)({\cal Q}) - {1
\over 2} \int_{\cal Q}^{{\cal R}^\prime} d \lambda \, \, \Box R
\nonumber \\
\mbox{} &=& e^{\Psi({\cal Q})} \, (l^a_\Sigma \nabla_a R)({\cal Q}) - {1
\over 2} \int_{\cal Q}^{{\cal R}^\prime} d \lambda \, \, \Box R,
\end{eqnarray}
where ${\cal R}^\prime$ is some point on $\Gamma$ between ${\cal Q}$
and ${\cal R}$.  Integrating once more with respect to $\lambda$
and using Eq.~(\ref{fla-figureofmerit}) yields
\begin{eqnarray}
\FL
|\sigma^{({\rm locally\ generated})}({\cal R})| &\le& {1 \over 48 \pi}
\Delta \lambda_{\rm max} e^{\Psi_{\rm max}} || l^a_\Sigma \nabla_a R||_\infty
\nonumber \\
\mbox{} && + {1 \over 96 \pi} (\Delta \lambda_{\rm max})^2 || \Box R ||_\infty,
\label{fla-proof1}
\end{eqnarray}
where $\Psi_{\rm max}$ is the maximum value of $\Psi$,
$|| l^a_\Sigma \nabla_a R||_\infty < \infty$ is the maximum value of
$l^a_\Sigma \nabla_a R$ on $\Sigma$, and $|| \Box R ||_\infty < \infty$
is the maximum value of $\Box R$ on $M$.  It follows from
Eq.~(\ref{fla-proof1}) that $\sigma^{({\rm locally\ generated})}({\cal
R})$ is bounded.

\subsection{Discussion of instability}
\label{fla-discussion}

As discussed in the Introduction, one can intuitively understand the
reason for the instability in the
following way.  The locally generated term in Eq.~(\ref{fla-basic1})
describes local particle creation and/or vacuum polarization effects
due to the background gravitational field.  [In general dynamic
spacetimes it is not meaningful to distinguish between particle
creation and vacuum polarization effects.]  Radiation propagating
parallel to the Cauchy horizon is generated all along the geodesic
$\Gamma$, and since the length of this geodesic is becoming infinite,
an infinite amount of radiation can be accumulated at the Cauchy
horizon.

Consider now the issue of under what conditions the divergence of
affine parameter length is a sufficient condition for a locally
created energy flux
instability of the Cauchy horizon.  First, as we discuss in
Sec.~\ref{fla-Misner}
below, there are spacetimes such as Misner space which are locally
flat so that the locally generated term (\ref{fla-figureofmerit}) vanishes
identically, but for which nevertheless the condition of divergence of
affine parameter length is satisfied.  Therefore, the
divergence of affine parameter length is not a sufficient condition
for {\it test field} locally created energy flux instabilities.
However, the example
of Misner is very special --- it is locally flat.  In general
spacetimes, one might imagine that if the curvature were very small or
vanishing near the Cauchy horizon in the background spacetime, then
the locally created energy flux instability might be present in second order
semiclassical perturbation if not in first order:  the first order
perturbation would give rise to some curvature near the Cauchy
horizon, and this curvature would then act as a source and give rise
to an divergence of the expected stress tensor at second order.
Therefore, we conjecture that the divergence of affine parameter
length is a {\it sufficient} condition for a full, nonlinear
instability of Cauchy horizons in dynamical, two dimensional
semiclassical theories with backreaction \cite{fla-note2D}.

\subsection{Explanation of instability as a ``delayed blueshift''
instability}.
\label{fla-delayedblueshift}

Our classification of Cauchy horizon instabilities is based on the
split (\ref{fla-generalsplit}) of the stress tensor into ``initial data''
and ``locally generated'' pieces.   However, this split depends on the
location of the initial data surface $\Sigma$.  An instability which
is a locally generated energy flux instability from the point of the
initial data surface $\Sigma$ can instead appear to be an
initial-data-related blueshift instability from the point of view of
some later surface $\Sigma^\prime$.  An example of this
behavior is given in Sec.~\ref{fla-RNdeS} below.  We now show that this
situation always occurs when the spacetime is blueshift stable but
semiclassically unstable: there is always some initial data surface
$\Sigma^\prime$ with respect to which the instability is a blueshift
instability.

The blueshift factor $\Psi({\cal R})$ defined in
Sec.~\ref{fla-preliminary} above depends not only on the point ${\cal
R}$ but also implicitly on the point ${\cal P}$ on the Cauchy horizon.
In this section we will write the blueshift factor as
$$
\Psi_{\cal P}({\cal R})
$$
to make the dependence on ${\cal P}$ explicit \cite{fla-noteBF}.
Consider now the
construction illustrated in Fig.~\ref{fla-fig2}.  Let ${\cal
R}^\prime$ be an arbitrary point on the geodesic $\Gamma$ between
${\cal R}$ and ${\cal Q}$, and let ${\cal R}_0^\prime$ and ${\cal
P}^\prime$ be corresponding points on the geodesic $\Gamma_0$ and on
the Cauchy horizon respectively, so that ${\cal R}_0^\prime$, ${\cal
R}^\prime$
and ${\cal P}^\prime$ all lie on a null geodesic parallel to ${\vec
l}$.  Let $\lambda_0$ be the affine parameter along the geodesic
$\Gamma_0$, so that ${\vec k} = d / d \lambda_0$ along $\Gamma_0$.
Then it is possible to show that the affine parameter length $\Delta
\lambda({\cal R})$ of the geodesic $\Gamma$ is given by the following
integral along the geodesic $\Gamma_0$:
\beq
\label{fla-db1}
\Delta \lambda({\cal R}) = \int_{\Gamma_0} d\lambda_0 \,\, {
e^{ \Psi_{\cal P}({\cal R})} \over
e^{ \Psi_{{\cal P}^\prime}({{\cal R}^\prime})} }.
\endeq
Now, if the total blueshift factor $\Psi_{\cal P}({\cal R})$ is
globally bounded, and if $\Delta \lambda({\cal R}) \to \infty$ as
${\cal R} \to {\cal P}$, it follows from Eq.~(\ref{fla-db1}) that the
quantity $e^{ \Psi_{{\cal P}^\prime}({{\cal R}^\prime})}$ cannot be
globally bounded below by a positive constant.  Therefore, for some
point ${\cal P}^\prime$ on  
the Cauchy horizon (passing to some conformal completion of the
spacetime if necessary), there is a diverging redshift, i.e.
$e^{ \Psi_{{\cal P}^\prime}({{\cal R}^\prime})} \to 0$ as ${\cal
R}^\prime \to {\cal P}^\prime$.

Consider now an ``initial data surface'' $\Sigma^\prime$ for which the
rightmost portion is a null geodesic parallel to ${\vec l}$ which
intersects the Cauchy horizon at ${\cal P}^\prime$.  Denote by
$\Psi^\prime_{\cal P}({\cal R})$ the
blueshift factor given by the construction of
Sec.~\ref{fla-preliminary}, with respect to the surface
$\Sigma^\prime$.  Then one has
\beq
\Psi^\prime_{\cal P}({\cal R}) = \Psi_{\cal P}({\cal R}) -
 \Psi_{{\cal P}^\prime}({{\cal R}^\prime}),
\endeq
which diverges as ${\cal R} \to {\cal P}$.  Therefore, the Cauchy
horizon is blueshift unstable with respect to the surface
$\Sigma^\prime$.  As explained in the Introduction, any energy flux
present on the initial surface $\Sigma$ will not give rise to a
divergence of the stress tensor on the Cauchy horizon, since the net
blueshift is finite.  By contrast, an energy flux which is created
locally near $\Sigma^\prime$ will suffer an infinite blueshift and
give rise an instability.

\subsection{Relevance to four dimensional spacetimes}
\label{fla-fourdimensions}

We now discuss the issue of to what extent this two dimensional
locally created energy flux instability mechanism is relevant to four
dimensional spacetimes.
Suppose one has a four dimensional spacetime on which
propagates a quantized scalar field ${\hat \Phi}$.  In
this context, the expected stress tensor is not
determined by its value on an initial surface $\Sigma$, so the split
(\ref{fla-generalsplit}) of the expected stress tensor in any state into a
``locally generated'' piece plus an ``initial data'' piece would not
seem to have an analog in four dimensions.  However, we now show that
this splitting, and in addition some of the features of the two dimensional
semiclassical theory which underly the instability mechanism, do
generalize to at least a certain class of four dimensional spacetimes.

Consider a four dimensional spacetime $(M,g_{ab})$ which, to the past
of some Cauchy surface $\Sigma$, is isometric to a portion of Minkowski
spacetime.  For any quantum state, let $G(x,y) = \langle {\hat
\Phi}(x) \, {\hat \Phi}(y) \rangle$ be that state's the two point
function, and let $G_0(x,y)$ be the two point function of the state
which to the past of $\Sigma$ is just the usual Minkowski vacuum.  Let
\beq
F(x,y) = G(x,y) - G_0(x,y),
\endeq
which is a smooth bisolution of the wave equation.  Then, as explained
in detail in Ref.~\cite{fla-anec}, the total expected stress tensor {\it
can} be written in a form analogous to Eq.~({\ref{fla-generalsplit}):
\beq
T_{ab} = T_{ab}^{({\rm initial\ data})} + T_{ab}^{({\rm locally\ generated})},
\label{fla-generalsplit4}
\endeq
where now

\begin{itemize}

\item The ``initial data'' piece is given by $T_{ab}^{({\rm
initial\ data})}  = {\rm Lim}_{y \to x} {\cal D}_{ab} F(x,y)$, where ${\cal
D}_{ab}$ is a second order differential operator.  This is the same
stress tensor as would be obtained in a purely {\it classical} theory
of a statistical ensemble of scalar fields, for which the stress
tensor is determined by the expected value $F(x,y) = \langle
\Phi(x) \Phi(y) \rangle$ of $\Phi(x) \Phi(y)$ with respect to the
classical ensemble.  Moreover, the quantity $F(x,y)$ is uniquely
determined (both classically and semiclassically) by its
initial data and derivatives on $\Sigma \times \Sigma$ from the
classical equations of motion \cite{fla-anec}.  Just as in two dimensions,
the main difference between the classical and
semiclassical theories is that the class of allowed initial data
(here, initial data for $F$ and its derivatives on $\Sigma \times
\Sigma$) is larger in the semiclassical case than in the classical
case.

\item The remaining ``locally generated'' piece $T_{ab}^{({\rm
locally\ generated})}$ of the stress tensor is the same for all
quantum states, and is just a unique functional of the spacetime
geometry, just as in two dimensions.   An explicit formula for this
piece of the stress tensor is known for the special case of metrics
that are linearized perturbations off Minkowski spacetime
\cite{fla-Horowitz80}.

\end{itemize}

These similarities between the two dimensional and four dimensional
semiclassical theories indicate that

\begin{itemize}

\item As in two dimensions, so also in four dimensions, one can probably
classify instabilities at Cauchy horizons into two types (i) ``blueshift
type'' instabilities that cause a divergence of the term
$T_{ab}^{({\rm initial\ data})}$ and that operate both classically and
semiclassically; and (ii) intrinsically quantum mechanical,
``locally created energy flux type'' instabilities that cause a
divergence of the term $T_{ab}^{({\rm locally\ generated})}$.

\item The property of divergence of affine parameter length {\it
might} be relevant to divergences of the term $T_{ab}^{({\rm
locally\ generated})}$ in four dimensions.  On the other hand, other
issues such as the focusing/defocusing along any geodesic, which enter
in four dimensions \cite{fla-Thorne93} but not in two, probably complicate
the situation.

\end{itemize}

In summary, it is an interesting open question whether or not any of
the ideas and results discussed here can be extended to four
dimensions; it seems conceivable that some of them could be.

\section{EXAMPLE SPACETIMES}
\label{fla-examplesec}

In this section, in order to illustrate and clarify the above
discussions and results, we examine several specific spacetimes ---
the two dimensional Reissner-N\"ordstrom-deSitter spacetime in
Sec.~\ref{fla-RNdeS}, general two dimensional spacetimes with closed
timelike curves in Sec.~\ref{fla-ctcs}, and Misner space in
Sec.~\ref{fla-Misner}.

\subsection{The two dimensional Reissner-N\"ordstrom-de Sitter
spacetime}
\label{fla-RNdeS}

The metric for a spherically symmetric black hole of mass $m$ and
charge $e$ in de-Sitter space can be written
as \cite{fla-Carter}
\beq
ds^2 = - f(r) dt^2 + f(r)^{-1} dr^2 + r^2 d\Omega^2,
\label{fla-bhmetric}
\endeq
where
\beq
f(r) = 1 - {2M \over r} + {e^2 \over r^2} - {1\over3} \Lambda r^2,
\endeq
and $\Lambda$ is the cosmological constant.  There are three positive
roots of the equation $f(r)=0$, which we will denote as $r_i$, $r_e$
and $r_c$, following Ref.~\cite{fla-MarkovicPoisson95}.  Here $r_i < r_e <
r_c$, and these three roots correspond to the inner or Cauchy horizon,
the event horizon, and the cosmological horizon, respectively (see
Fig.~\ref{fla-fig3}).  We also define $\kappa_j = |f'(r_j)|/2$, which
is the surface gravity at the $j$th horizon, for $r_j = r_i$, $r_e$ or
$r_c$.  We will discuss for the most part only the two dimensional
version of the spacetime in which the last term in Eq.~(\ref{fla-bhmetric}) is
omitted.

In Ref.~\cite{fla-BradyPoisson92}, Brady and Poisson showed that the Cauchy
horizon is classically test field stable in the region $\kappa_i < \kappa_c$
of parameter space, when the matter model is taken to consist of an
infalling null fluid (or equivalently a minimally coupled scalar
field in the two dimensional context), and they argued that the
stability result should also hold for more realistic matter models and
for gravitational perturbations.  In
Ref.~\cite{fla-MarkovicPoisson95},
Markovi\'{c} and Poisson showed that the two dimensional spacetime is
semiclassically unstable, by showing that the stress tensor diverges
on the Cauchy horizon in the Markovi\'{c}-Unruh state
\cite{fla-MarkovicUnruh91}, and in Ref.~\cite{fla-Poisson97} Poisson extended
this result to show that the
stress tensor must diverge on the Cauchy horizon in {\it any} state
which is regular on the cosmological and event horizons.  (See also
earlier calculations by Davies and Moss \cite{fla-DaviesMoss} which
indicated that the Cauchy horizon is unstable, and see
Ref.~\cite{fla-Chambers97} for a detailed overview of this
literature).

In this section, we will focus attention on some of the above results,
and for the most part simply verify that they are reproduced by the
general formalism of Secs.~\ref{fla-foundations} -- \ref{fla-affinesec}.  We
will also show that in the integral (\ref{fla-figureofmerit}), both the
integrand $l^a \nabla_a R$ and the total affine parameter length
diverge, as claimed in Sec.~\ref{fla-affinesec}.

{\vskip 1cm}

{\psfig{file=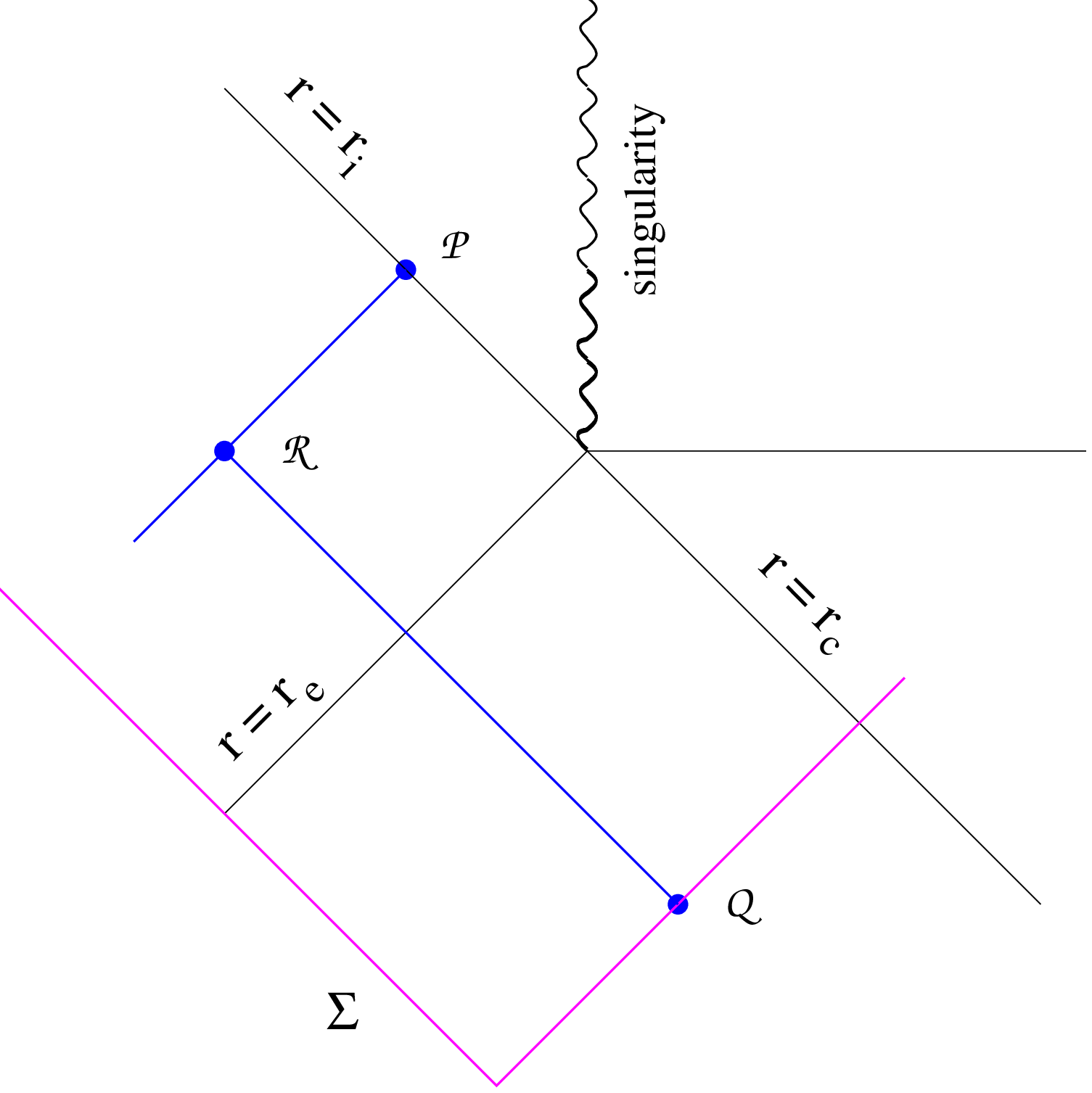,height=8cm,width=8cm,angle=-90}}
{\vskip 0.5cm}
\begin{figure}
\caption{
The Penrose diagram for a portion of the two dimensional
Reissner-N\"ordstrom-deSitter spacetime which evolves from data
specified on the initial Cauchy surface $\Sigma$.  The lines marked
$r=r_c$, $r=r_e$ and $r=r_i$ are the cosmological, event and
inner or Cauchy horizons respectively.  Also shown are the geodesics
$\Gamma$ and $\Lambda$ and the points ${\cal P}$, ${\cal R}$ and
${\cal Q}$ of the general construction of Sec.~\ref{fla-nullbasissec}.
}
\label{fla-fig3}
\end{figure}

{\vskip 0.5cm}

We take the initial surface to consist of two null half-lines, as
illustrated in Fig.~\ref{fla-fig3}, where the rightmost half-line is
outside the event horizon.  The null basis $\{{\vec k}, {\vec l} \}$
must have the form
\begin{eqnarray}
{\vec k} &=& \alpha \left[ {1 \over f} {\partial \over \partial t} -
{\partial \over \partial r} \right] \nonumber \\
\mbox{} {\vec l} &=& {f \over 2 \alpha} \left[ {1 \over f} {\partial
\over \partial t} + {\partial \over \partial r} \right],
\label{fla-cc1}
\end{eqnarray}
where $\alpha$ is some positive function on spacetime, since the
vectors ${\vec k}$, ${\vec l}$ are future directed and null.
Now the two dimensional version of the metric (\ref{fla-bhmetric}) can be
written in the usual way as $-f
(dt^2 - dr_*^2)$, and since in any spacetime of the form
(\ref{fla-conformalcoords}) the vector fields $e^{-\sigma} \partial_u$ and
$e^{-\sigma} \partial_v$ are geodesic, it follows that the vector
fields $\pm f^{-1} \partial_t \pm \partial_r$ are locally geodesic.
Therefore since ${\vec l}$ is geodesic along the geodesic $\Lambda$ in
Fig.~\ref{fla-fig3}, it follows from Eq.~(\ref{fla-cc1}) that $f / \alpha$
must be constant along $\Lambda$, and without loss of generality we
can take this constant to be $-1$:
\beq
\alpha({\cal R}) = - f({\cal R}), \,\,\,\, {\cal R} {\mbox \ \ {\rm on}\ }
\Lambda.
\label{fla-cc2}
\endeq
Similarly, ${\vec k}$ is geodesic along $\Gamma$, so $\alpha$ is
constant along $\Gamma$,
\beq
\alpha({\cal R}^\prime) = \alpha({\cal R})
\label{fla-cc3}
\endeq
where ${\cal R}^\prime$ is any point along $\Gamma$.  Combining
Eqs.~(\ref{fla-cc1}) -- (\ref{fla-cc3}) we find that the integrand in
Eq.~(\ref{fla-figureofmerit}) is given by
\beq
(l^a \nabla_a R)({\cal R}^\prime) = - {f({\cal R}^\prime) \over 2
f({\cal R}) } R'(r),
\label{fla-bhintegrand}
\endeq
where $R = R(r)$ is the Ricci scalar which depends only on $r$.

Next, from Eq.~(\ref{fla-cc1}) we find that the relationship between the
affine parameter $\lambda$ along $\Gamma$ and the coordinate $r$ is
given by $d \lambda = - d r / \alpha$, which from Eq.~(\ref{fla-cc3})
gives
\beq
\lambda = {1 \over f({\cal R})} \left[ r + {\rm const}\right].
\label{fla-bhaffine}
\endeq
Combining Eqs.~(\ref{fla-figureofmerit}), (\ref{fla-bhintegrand}) and
(\ref{fla-bhaffine}) now gives
\beq
\FL
\sigma^{({\rm locally\ generated})}({\cal R}) =
{1 \over 96 \pi f({\cal R})^2}
\int_{r({\cal R})}^{r[{\cal Q}({\cal R})]} \, dr f(r) R'(r).
\endeq
Here as before ${\cal Q} = {\cal Q}({\cal R})$ denotes the unique
point ${\cal Q}$ on the initial surface $\Sigma$ determined by ${\cal
R}$ according to the construction of Sec.~\ref{fla-nullbasissec}.
Using the formula $R(r) = - f''(r)$ and integrating by parts gives
\beq
\int_{r({\cal R})}^{r[{\cal Q}({\cal R})]} \, dr \, f(r) R'(r) = \left[
{1\over2} (f^\prime)^2 - f f'' \right]^{r[{\cal Q}({\cal R})]}_{r({\cal R})}.
\endeq
Finally, using that fact that as ${\cal R} \to
{\cal P}$, the limits of integration behave as $r({\cal R}) \to
r_i$, $r[{\cal Q}({\cal R})] \to r_c$ and that $f(r_i) = f(r_c) =0$
gives
\beq
\FL
\sigma^{({\rm locally\ generated})}({\cal R}) \approx ({\rm const}) \,\,
{1 \over f({\cal R})^2} \left[ \kappa_c^2 - \kappa_i^2 \right].
\label{fla-bhfinalans}
\endeq
Equation (\ref{fla-bhfinalans}) reproduces the result of
Refs.~\cite{fla-MarkovicPoisson95,fla-Poisson97} that $\sigma$ is divergent on
the Cauchy horizon in the classically stable region $\kappa_c >
\kappa_i$ of parameter space, since $f({\cal R}) \to 0$ as ${\cal R}
\to {\cal P}$.  The agreement with Ref.~\cite{fla-MarkovicPoisson95} can
be made more explicit by noting that in this limit $f({\cal R})
\approx ({\rm const}) \, \exp[ - \kappa_i v({\cal R})]$, where $v$ is
the advanced time coordinate $t + r_*$ which goes to infinity as
${\cal R} \to {\cal P}$ \cite{fla-BradyPoisson92}.

Several points should be noted about the above derivation.  First,
from Eq. (\ref{fla-bhaffine}) it can be seen that
the affine parameter length is divergent:
\beq
\Delta \lambda({\cal R}) \approx { 1 \over f({\cal R}) } (r_c - r_i)
\approx e^{\kappa_i v} \, (r_c - r_i);
\endeq
and that the integrand (\ref{fla-bhintegrand}) itself also has an overall
factor of $1/f({\cal R})$ which is divergent as ${\cal R} \to {\cal
P}$ \cite{fla-note6}.  Second, the instability result applies to {\it all}
quantum
states for which the initial data on $\Sigma$ is regular.  The
quantity $\sigma({\cal R})$
will be a sum of the divergent, locally generated piece
(\ref{fla-bhfinalans}), together with the piece (\ref{fla-blueshiftans}) which
depends on the initial data and which will be finite as the blueshift
factor $e^\Psi$ is finite.  [It is easy to verify that up to an
overall constant factor the blueshift factor is
\begin{eqnarray}
e^{\Psi({\cal R})} &=& { f[{\cal Q}({\cal R})] \over f({\cal R}) }
\nonumber \\
\mbox{} &\approx& {e^{-\kappa_c v({\cal R})} \over e^{-\kappa_i
v({\cal R})}} \ \ \ \ \ \mbox{{\rm as}} \ \  {\cal R} \to {\cal P}.
\end{eqnarray}
This is finite for $\kappa_c > \kappa_i$, as first noted in
Ref.~\cite{fla-BradyPoisson92}].  Third, the derivation allows us to
understand why the divergent piece of the quantity $\sigma({\cal R})$
in semiclassical analyses does not contain a cosmological redshift
factor $e^{- 2 \kappa_c v}$ as it does in classical analyses
\cite{fla-MarkovicPoisson95}.  The
reason is that the radiation giving rise to the divergence
(\ref{fla-bhfinalans}) is produced in the vicinity of the event horizon
and propagates from there inward to the Cauchy horizon,
and thus is not affected by the redshift factor describing propagation
from the cosmological horizon to the event horizon.

Note that it is straightforward to show that
in the classically stable region $\kappa_c > \kappa_i$ of parameter
space, the blueshift factor $e^{\Psi({\cal R})}$ is globally bounded,
not just inside the black hole but also along the event horizon.
Thus, the class of spacetimes to which the general result of
Sec.~\ref{fla-necessarycondt} applies is not empty.

We have explained the above instability as a locally created energy
flux instability and not as a blueshift instability.  However, it
can also be thought of as a ``delayed blueshift'' instability, as
discussed in Sec.~\ref{fla-delayedblueshift} above.
If one chooses the right-most portion of the
initial data surface $\Sigma$ to lie on the event horizon, then the
corresponding blueshift factor given by the construction of
Sec.~\ref{fla-preliminary} is proportional to $e^{2 \kappa_i v}$ and
so is unbounded.  In classical analyses, the initial data on the event
horizon must fall off like $e^{-2 \kappa_c v}$ (due to having
propagated inward from some initial surface outside the event
horizon), thus giving rise to a finite value of $\sigma$ on the Cauchy
horizon.  However, in a semiclassical analysis, things are different.
Consider an ingoing mode of the quantum field whose mode function is
concentrated near some geodesic that is very close to the Cauchy
horizon.  Any ingoing quanta in this mode from outside the black hole
experience first a large redshift (from the outside to the event
horizon), then a large blueshift (from the event horizon to the
interior) which is smaller than the redshift.  The crucial feature in
the semiclassical analysis is that ingoing quanta
are created in the vicinity of the event horizon; these propagate
inwards and suffer only the large blueshift.  Thus, the
``initial data'' for the stress tensor on the event horizon need not
fall off like $e^{-2 \kappa_c v}$ in the semiclassical theory.
Correspondingly, near the Cauchy horizon $\sigma$ diverges like $e^{2
\kappa_i v}$ by the usual blueshift effect.

Thus, the distinction between blueshift and locally created energy
flux instabilities is dependent on the choice
of location of the initial data surface, as discussed in
Sec.~\ref{fla-delayedblueshift} above

\subsection{General spacetimes with closed timelike curves}
\label{fla-ctcs}

A special case of the general class of spacetimes described in
Sec.~\ref{fla-class} above is when the spacetime $(M^\prime,
g_{ab}^\prime)$ contains closed timelike curves, so that the Cauchy
horizon $H^+(\Sigma)$ is also a chronology horizon.  A simple example
of such a spacetime is given by Yurtsever
\cite{fla-Yurtsever91}, in which the lightcones on a cylinder ``tip over''
to produce a closed null geodesic around the cylinder \cite{fla-noteCNG}.
This closed
null geodesic coincides with the chronology horizon.  [In the four
dimensional context, closed null geodesics or ``fountains'' will form
a small subset of the full chronology horizon \cite{fla-Thorne93}.]

In Ref.~\cite{fla-Hawking92}, Hawking gives a general argument, in the
context of chronology horizons in four dimensional spacetimes which
are compactly generated, that the total affine parameter length of
all fountains should be infinite in the past direction (although
generically finite in the future direction):  If ${\vec k}$ is a future
directed, null vector which is tangent to the geodesic at some point,
then the total holonomy around the fountain in the future direction
will map ${\vec k}$ to $e^h {\vec k}$ for some constant $h$.  Hawking shows
that
if $h<0$, then there must exist a closed timelike curve to the past of
the chronology horizon, which is a contradiction.  Therefore $h \ge 0$
and so the the total affine
parameter length in the past direction of the closed null geodesic is
proportional to
\beq
\sum_{n=0}^\infty e^{n h} = \infty,
\endeq
(although the total affine parameter length in the future direction is
proportional to $\sum e^{- n h} < \infty$ as long as $h \ne 0$).
It can be checked that this argument applies equally well to two
dimensional spacetimes.  In the two dimensional context,
the generator of the Cauchy horizon through any point ${\cal P}$ will
therefore have infinite affine parameter length in the past direction,
and it follows from the argument of appendix \ref{fla-generatorappendix} that
all such spacetimes satisfy the property (\ref{fla-DAPL}) of divergence of
affine parameter length.

However, it does not follow that all such spacetimes are subject to
the locally created energy flux instability.  We now turn to a
specific spacetime containing closed timelike curves which illustrates
this point.

\subsection{Misner space}
\label{fla-Misner}

Misner space is a well-known locally flat spacetime with topology $S^1
\times R$ \cite{fla-Misner1,fla-HawkingEllis73}.  As is well known,
Misner space is both
classically and semiclassically unstable
\cite{fla-HiscockKonkowski82}.  These instability properties of
the spacetime are reviewed in Ref.~\cite{fla-Misner2}.
In this section we show how the construction of
Secs.~\ref{fla-blueshiftsec} and \ref{fla-affinesec} applies to Misner space.
The construction reproduces the well-known fact that Misner space
suffers from the blueshift instability, which explains both the
classical and semiclassical instabilities.  We also show that Misner
space does {\it not} suffer from the locally created energy flux
instability, despite the fact that the spacetime does
satisfy the property of divergence of affine parameter length.

Misner space can be constructed as follows \cite{fla-Misner2}.  In two
dimensional Minkowski spacetime with coordinates $(t,x)$, identify the
worldline $x=0$ with the worldline $x = L - \beta t$, for some $L >
0$, $\beta > 0$ (see Fig.~\ref{fla-fig4}).  Let $\tau$ be the proper
time measured by a clock on
one of these world lines; $\tau = t$ for the leftmost worldline,
whereas $\tau = t/ \gamma$ for the rightmost worldline, where $\gamma$
is the usual special relativistic time dilation factor.  An observer
crossing the rightmost worldline at coordinate time $t=t_0$ therefore
emerges from the leftmost line at coordinate time $t = t_0/\gamma$.
Hence, at large $t$, there will be closed timelike curves in the
spacetime, after the closed null geodesic marked in
Fig.~\ref{fla-fig4}.  This closed null geodesic is the Cauchy
horizon or chronology horizon of Misner space.

Consider now how the constructions of Secs.~\ref{fla-nullbasissec} and
\ref{fla-preliminary} apply to Misner space.  First, note that if any
vector is parallel transported through the leftmost worldline to
emerge from the rightmost worldline, it undergoes the boost given by
$\partial_u \to e^\xi \partial_u$, $\partial_v \to e^{-\xi}
\partial_v$, where $\cosh \xi = \gamma$ and $v = t+x$, $u=t-x$.
{}From the diagram of Misner space shown in Fig.~\ref{fla-fig4},
we can fairly easily see that (i) If we take the initial surface
$\Sigma$ to be the surface $t=0$, then $\Sigma$ has the topology of a
circle.  The basis $\{{\vec k}_\Sigma, {\vec l}_\Sigma\}$ obtained by
parallel
transport along $\Sigma$ will be discontinuous at some point $Q_0$,
because of the overall boost when passing through the identified
worldlines.  (ii) The null geodesic $\Gamma$ through a point ${\cal
R}$ near the Cauchy horizon passes through the identified worldlines
some number $n$ of times before it reaches the initial surface, where
$n$ grows without limit as ${\cal R} \to {\cal P}$.
Thus, the initial point ${\cal Q}({\cal R})$ circles around and
around $\Sigma$, as claimed in Sec.~\ref{fla-gg} above.  (iii)
When one parallel transports the vector ${\vec k}$ along $\Gamma$ to
$\Sigma$, it is boosted $n$ times, and thus the redshift factor is
\beq
e^{\Psi({\cal R})} = e^{n \xi},
\endeq
which diverges as ${\cal R} \to {\cal P}$.  This redshift factor
changes discontinuously by a factor of $e^\xi$ every time ${\cal
Q}({\cal R})$ crosses ${\cal Q}_0$ on $\Sigma$, as the basis $\{ {\vec
k}_\Sigma, {\vec l}_\Sigma \}$ is
discontinuous there.  (iv) The total affine
parameter length $\Delta \lambda({\cal R})$ of $\Gamma$ also diverges
as ${\cal R} \to {\cal P}$.  [This also follows from the general
argument of Sec.~\ref{fla-ctcs} above].

{\vskip 1cm}

{\psfig{file=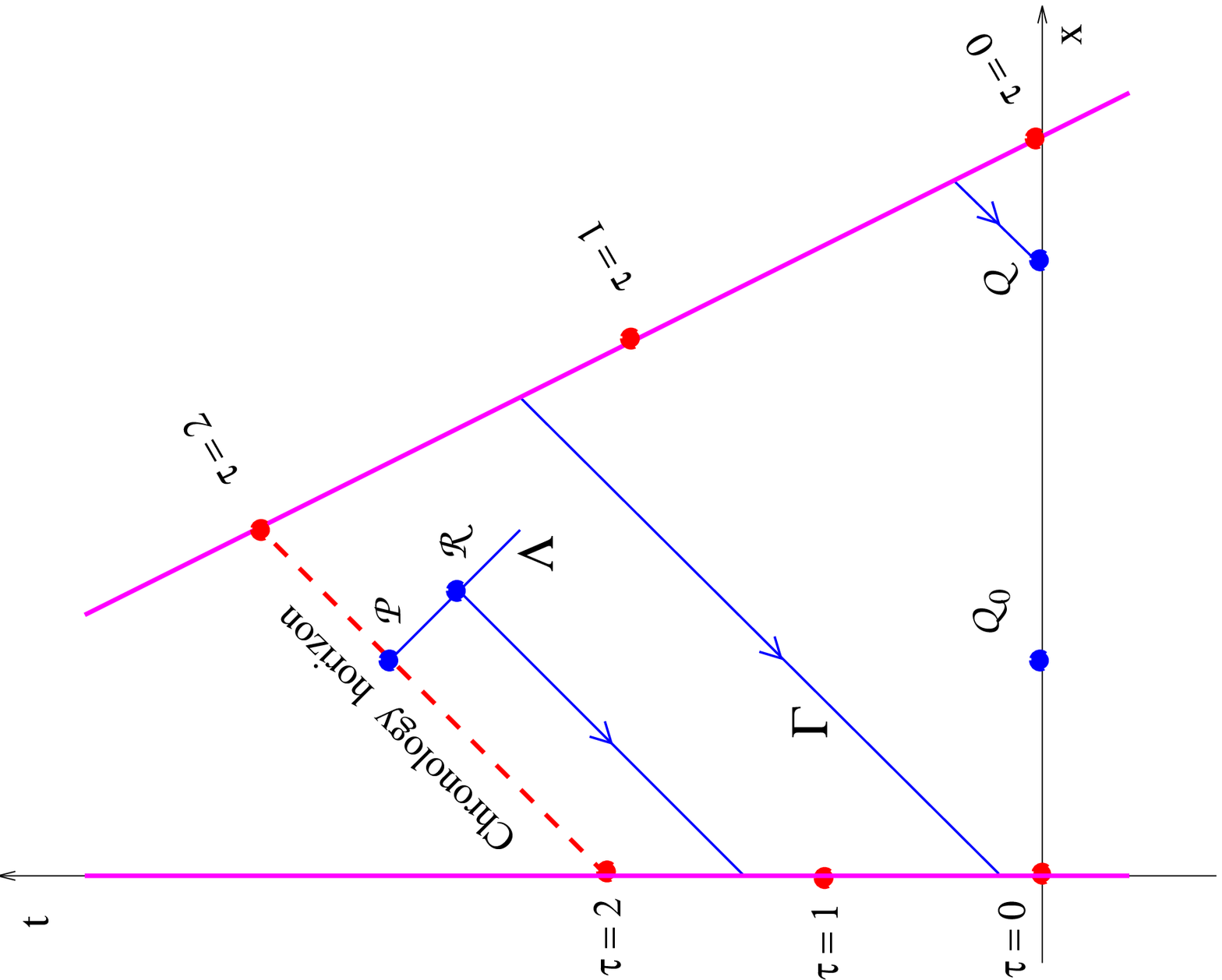,height=10cm,width=8cm,angle=-90}}
{\vskip 0.5cm}
\begin{figure}
\caption{A spacetime diagram of Misner space (adapted from
Ref.~[53]).  The two thick black lines
are identified.  Shown are a representative point ${\cal P}$ on the
chronology horizon, the null geodesic $\Lambda$ emanating from ${\cal
P}$, a point ${\cal R}$ on $\Lambda$ near ${\cal P}$, and the null
geodesic $\Gamma$ joining ${\cal R}$ to the initial surface $t=0$ at
the point ${\cal Q}$.  It can be seen that the spacetime satisfies the
property of divergence of affine parameter length (as must all
spacetimes with closed timelike curves) since the number of times the
geodesic $\Gamma$ winds around the spacetime diverges as ${\cal R} \to
{\cal P}$.  Moreover the blueshift factor $e^{\Psi({\cal R})}$, which
is obtained from the Lorentz transformation got by parallel
transporting along $\Gamma$ between ${\cal Q}$ and ${\cal R}$, also
diverges as ${\cal R} \to {\cal P}$, because all vectors are boosted
when they pass through the identified worldlines.}
\label{fla-fig4}
\end{figure}

Since the blueshift factor diverges, the spacetime is classically and
quantum mechanically unstable, from the arguments of
Sec.~\ref{fla-blueshiftsec} above.  In the classical case, this means that
the stress tensor diverges on the Cauchy horizon for generic initial
data of the scalar field \cite{fla-MisnerUnstable}.   Of course, for the
exceptional case of vanishing initial data on $\Sigma$, the stress
tensor vanishes identically.  Nevertheless, one still regards the
Cauchy horizon as being unstable as vanishing initial data is not
generic or stable in any physical sense.

The situation is very similar in the semiclassical theory.
Originally it was conjectured in
Ref.~\cite{fla-Hawking92} that the stress
tensor must diverge on the Cauchy horizon for {\it all} quantum
states.  The argument of Sec.~\ref{fla-blueshiftsec} shows that the stress
tensor must diverge unless the initial data $\sigma_\Sigma$ on
$\Sigma$ is vanishing; thus, the conjecture was effectively that there
is no state on Misner space for which the expected stress tensor vanishes
identically on the initial surface.  However, it is now known that
there are quantum states on both two dimensional and four dimensional
Misner space for which the stress tensor does vanish identically
\cite{fla-Sushkov97,fla-Krasnikov96,fla-Kay97}.  Nevertheless, states
for which
the initial data for the stress tensor is vanishing are clearly in
some sense non-generic, and therefore one is still justified in
regarding Misner space as unstable in the semiclassical theory, just
as in the classical theory.

Finally, we note that the locally created energy flux
instability mechanism does not operate in Misner space, since it is
locally flat and so the integrand in Eq.~(\ref{fla-figureofmerit}) is
identically vanishing.  This does not contradict the fact that the
condition (\ref{fla-DAPL}) is satisfied by Misner space, since we have
only shown that the condition (\ref{fla-DAPL})
is a necessary condition for the instability, and
not a sufficient condition.  Note, however, that all spacetimes which
are ``close'' to Misner space but which differ from it by having an
arbitrarily small amount of spacetime curvature on the Cauchy horizon
should suffer from the locally created energy flux instability.  See
also the related discussion in Sec.~\ref{fla-discussion} above.

\section{CONCLUSION}
\label{fla-conclusions}

We have shown that in the context of semiclassical gravity in two
dimensions, there are two different types of instabilities of Cauchy
horizons.  The first is a divergence of the piece of the stress tensor
which is determined by the initial data; it operates classically as
well as semiclassically, and is just the well-understood blueshift
instability.  The second, which we call the locally created energy
flux instability, is
a divergence of the locally generated piece of the expected stress
tensor.  This instability is characterized by the divergence of the
affine parameter lengths of null geodesics parallel to and close to
the Cauchy horizon.

It is natural to conjecture that for all Cauchy horizons in two
dimensional spacetimes without singularities, one or other of these
two instability mechanism always applies, if not in linear
perturbation theory, then at least in nonlinear analyses.

\acknowledgements

It is a pleasure to thank Lior Burko, Amos Ori and Liz Youdim for
organizing the very enjoyable workshop on ``The Internal Structure of
Black Holes and Spacetime Singularities'' at the Technion, Haifa, and
for the 
generous hospitality offered during our stay there.  I also wish to
thank Patrick Brady, Chris Chambers, Ian Moss, Amos Ori, Eric Poisson,
Kip Thorne, and Robert Wald for helpful discussions on Cauchy horizon
stability issues.  This research was
supported in part by NSF grant PHY 9722189 and by a Sloan Foundation
fellowship.

\appendix
\section{Diverging affine parameter length condition follows from
Cauchy horizon generator being of infinite affine parameter length in
the past}
\label{fla-generatorappendix}

In this appendix we show that if the generator of the Cauchy horizon
through the point ${\cal P}$ has infinite affine parameter length in
the past, then the condition (\ref{fla-DAPL}) must be satisfied by the
spacetime.

Let $T M^\prime$ be the tangent bundle over the spacetime
$(M^\prime,g_{ab}^\prime)$, and let $$\exp : D \subset T M^\prime \to T
M^\prime$$ be the exponential map, where the domain of definition $D$
of the exponential map is an open subset of $T M^\prime$.
[If $(M^\prime,g_{ab}^\prime)$ were geodesically complete we would
have $D = T M^\prime$].  In other
words, for any point ${\cal B}$ in $M^\prime$ and any vector ${\vec
v}$ at ${\cal B}$, $\exp[{\cal B},{\vec v}]$ will be the pair $({\cal
B}^\prime, {\vec v}^\prime)$ in $TM^\prime$ such that the geodesic
starting at ${\cal B}$ with initial tangent ${\vec v}$ reaches
${\cal B}^\prime$ after one unit of parameter length, and such that
${\vec v}^\prime$ is the tangent to the geodesic at ${\cal B}^\prime$.

Let $U_+ = M^\prime - {\overline {D^-(\Sigma)}}$,
the complement of the closure of the past domain of dependence of
$\Sigma$, and let $W_+ \subset TM^\prime$ be the tangent bundle over
$U_+$.  It is clear that the set $W_+$ is open in
$TM^\prime$.  Consider now the construction outlined in
Sec.~\ref{fla-nullbasissec}. It can be seen that, for any $l > 0$ and for
any ${\cal R}$ on $\Lambda$,
\beq
\exp[ {\cal R}, - l \, {\vec k}({\cal R}) ] \in W_+ \ \ \ \ \
{\rm implies} \ \ \ \ \Delta \lambda({\cal R}) \ge l,
\endeq
by the definition of $\Delta \lambda({\cal R})$ given in
Sec.~\ref{fla-affinesec} above and using the fact that $\Sigma$ is a
Cauchy surface for $(M,g_{ab})$.

Suppose now that the affine parameter length towards the past of the
generator of the Cauchy horizon through ${\cal P}$, normalized with
respect to ${\vec k}({\cal P})$, is greater than
some number $\beta$.  It follows that $[{\cal P}, - \beta \, {\vec
k}({\cal P})]$ lies in the domain of definition $D$ of the exponential
map.  Also it follows that $\exp[{\cal P}, - \beta \, {\vec
k}({\cal P})]$ lies in $W_+$, and thus by continuity of the
exponential map there is some open neighborhood $U$ of $[{\cal P},
- \beta \,{\vec k}({\cal P})]$ in $D$ whose image under the
exponential map lies in $W_+$.  Hence there is some neighborhood $V$
of ${\cal P}$ in $M^\prime$ such that for all points ${\cal R}$ on
$\Lambda$ and in $V$, $\exp[{\cal R}, - \beta {\vec k}({\cal R})]$ is
defined and lies in $W_+$ and so $\Delta \lambda ({\cal R}) \ge
\beta$.  Since this is true for all $\beta$ the result follows.

Note that the converse of this result if not true:  if the diverging
affine parameter condition holds for a given point ${\cal P}$ on the
Cauchy horizon, it does not follow that the generator of the Cauchy
horizon through ${\cal P}$ has infinite affine parameter length in the
past.  This is because one always has the freedom to redefine the
spacetime $(M^\prime,g_{ab}^\prime)$ by excising points on the Cauchy
horizon.

We have chosen to express the condition in terms of the
behavior of the spacetime $(M,g_{ab})$ [divergence of affine
parameter length of null geodesics parallel to and close to the Cauchy
horizon] rather than the behavior of the larger spacetime $(M^\prime,
g_{ab}^\prime)$ [generator of Cauchy horizon having infinite affine
parameter length in the past] in order that the condition be
manifestly independent of which extension we pick.

\newpage
\onecolumn
\begin{table}
\caption{A table contrasting the two different instability mechanisms
discussed in this contribution, in the context of two dimensional
spacetimes: the {\it blueshift instability} characterized by the
divergence on the Cauchy horizon of the ``initial data'' piece of the
expected stress tensor, and what we call the {\it locally created energy flux
instability} mechanism, which is characterized by the divergence on
the Cauchy
horizon of the ``locally generated'' piece of the expected stress
tensor.
\label{fla-table1}}
\begin{tabular}{lll}
\mbox{}&\mbox{Blueshift instability}&\mbox{Locally created energy flux
instability}\\
\tableline
\ &\ &\ \\
\mbox{Applies classically?}&\mbox{Yes}&\mbox{No}\\
\ &\ &\ \\
\mbox{Applies semiclassically?}&\mbox{Yes}&\mbox{Yes}\\
\ &\ &\ \\
\mbox{Necessary condition}&\mbox{Divergence of
holonomy}&\mbox{Divergence of affine parameter length}\\
\mbox{\ for instability}&&\mbox{}\\
\ &\ &\ \\
\mbox{Sufficient condition}&\mbox{Divergence of
holonomy}&\mbox{Divergence of affine parameter length is}\\
\mbox{\ for instability}&\mbox{\ (with some conditions on initial
data)}&\mbox{\ probably a sufficient condition for nonlinear}\\
\mbox{}&\mbox{}&\mbox{\ instability, but is not for linear
instability}\\
\ &\ &\ \\
\mbox{Spacetimes in
which}&\mbox{Reissner-N\"ordstrom}&\mbox{some Reissner-N\"ordstrom-deSitter}\\
\mbox{\ applies}&\mbox{Misner
space}&\mbox{ spacetimes}\\
\ &\ &\ \\
\mbox{Spacetimes in
which}&\mbox{some Reissner-N\"ordstrom-deSitter
spacetimes}&\mbox{Misner space: linear instability does not}\\
\mbox{\ does not apply}&\mbox{}&\mbox{\ apply}\\
\tableline
\end{tabular}
\end{table}


\begin{references}
\bibitem{fla-notecomplete}
	We assume that the initial surface is a complete
	Riemannian manifold to exclude examples, such as the region
	$|u|<1$, $|v|<1$ of two dimensional Minkowski spacetime, where
	the predicted spacetime is incomplete simply because the
	initial data is incomplete.
\bibitem{fla-notesecond}
	Of course, it could also happen that one can have maximal
	Cauchy evolutions with both curvature singularities and Cauchy
	horizons.  However, this possibility is not so disturbing.
	Suppose we define the ``classical region'' of spacetime to
	consist of all points ${\cal P}$ for which the curvature is
	sub-Planckian everywhere to the past of ${\cal P}$; then if
	the Cauchy horizon is outside the ``classical region'' there is less
	cause for concern, as one would not believe the predictions of the
	classical theory outside the classical region anyway.
\bibitem{fla-noteterminology}
	By classical general relativity we mean that both the metric
	and matter are classical, as opposed to semiclassical general
	relativity.
\bibitem{fla-noteCTC}
	Ref.~\cite{fla-KimThorne91} showed that the expected stress tensor
	diverges on the Cauchy horizon for quantum states for which the two
	point function is dominated by a series of singular
	contributions which can be calculated using geometric optics.
	Recently,
	examples have been recently found of states of automorphic
	fields on four dimensional spacetimes with closed timelike curves
	for which the stress tensor does not diverge on the chronology
	horizon \cite{fla-Sushkov97}, and similar examples have been found
	for massless scalar fields in two dimensions \cite{fla-Krasnikov96}.
	Thus, the above property of two point functions used in
	Ref.~\cite{fla-KimThorne91} must not be valid for all states;
	however, it is presumably valid for generic states and thus
	the stress tensor should diverge on the chronology horizon for
	generic states.
\bibitem{fla-noteCTC1}
	It follows from our analyses that the instabilities of
	chronology horizons in two dimensions (and probably also in
	four dimensions) can be split into two categories:
	(i) blueshift instabilities, which
	apply both classically and semiclassically, as in Misner space, and
	(ii) locally created energy flux instabilities, which can be
	understood as the ``piling up
	of vacuum fluctuations'', as explained in Ref.~\cite{fla-Thorne93}.
	For locally created energy flux instabilities, a crucial
	feature of this pile-up is that each time a wavepacket of a particular
	mode of a quantum field propagates around an almost-closed null curve
	near the chronology
	horizon, the total expected value of propagating energy is augmented
	by a small amount.  [See Eq.~(\ref{fla-figureofmerit}) above and
	associated discussion.]  The instability cannot be explained just from
	the fact that the wavepackets from each successive trip around
	almost-closed null curves tend to overlap more and more as the horizon is
	approached; that overlap is also present classically.
\bibitem{fla-notebasis}
	The resulting basis can be expressed as
	$$
	{\vec l}(u,v) = e^{-\sigma(u_{\cal P},v)} \, {\partial \over
	\partial v}
	$$
	and
	$$
	{\vec k}(u,v) = e^{\sigma(u_{\cal P},v) - \sigma(u,v)} \,
	{\partial \over \partial u},
	$$
	for any double null coordinates $(u,v)$ for which the metric
	takes the form (\ref{fla-conformalcoords}) and which are regular
	at ${\cal P}$, where $u_{\cal P}$ is the value of the $u$
	coordinate at ${\cal P}$.
\bibitem{fla-note1}
	One might worry that if the behavior of, say, $\rho$ near
	${\cal P}$ were very direction dependent, then it might happen
	that $\rho[x^a(\lambda)]$ would diverge for some curves
	$x^a(\lambda)$ while nevertheless remaining always bounded for all
	timelike curves of bounded proper acceleration.
	This possibility is excluded by the equation of motion
	(\ref{fla-basic}) for the stress tensor: since the geometry is
	regular near ${\cal P}$, in a sufficiently small neighborhood
	of ${\cal P}$ the effect of the source term will be
	unimportant and to a good approximation $\rho$ will depend on
	only one of the two null coordinates $u$ and $v$.
\bibitem{fla-noteInitialData}
	Specifically, consider the special case where $\Sigma$
	consists of two intersecting null lines as illustrated in
	Fig.~\ref{fla-fig1}.  Then, the free data on one of these two
	half-lines consists of the function $\sigma(\lambda) \equiv
	T_{ab} (\partial / \partial \lambda)^a (\partial /
	\partial \lambda)^b$, where $\lambda$ is an affine parameter
	along the line.  In the classical theory, $\sigma(\lambda)$
	can be any non-negative smooth function.  In the semiclassical
	theory in the special case of two dimensional Minkowski
	spacetime, it follows from the analyses of Ref.~\cite{fla-QIpaper}
	that the class of allowed functions $\sigma(\lambda)$ consists
	of those smooth functions which satisfy
	$$
	\int d\lambda \, f(\lambda) \, \sigma(\lambda) \, \ge \,
	- {1 \over 48 \pi} \int d\lambda  \, {f'(\lambda)^2 \over
	f(\lambda)},
	$$
	for all smooth non-negative functions $f(\lambda)$ [where
	$f^\prime(\lambda)^2/f(\lambda)$ is interpreted to mean zero
	when $f(\lambda)=0$].  This
	characterization also applies to spacetimes which are globally
	conformal to two dimensional Minkowski spacetime, as can be
	seen from Eq.~(6.133) of Ref.~\cite{fla-BirrellDavies82}.
\bibitem{fla-notquite}
	This is not quite true as one could envisage a situation where
	the ``locally generated'' and ``initial data'' terms in
	Eq.~(\ref{fla-generalsplit}) both diverge but where they
	cancel each other
	out so that that the sum is finite.  We ignore this
	possibility here.  See however Ref.~\cite{fla-DaviesMoss} for a
	scenario where such a cancelation could occur.
\bibitem{fla-example2}
	An example of a spacetime where this occurs is the following:
	Let $M_0$ be the region $|u| < 1$, $|v|<1$ of Minkowski
	spacetime, where $u$ and $v$ are the usual null coordinates,
	and let $M$ be $M_0 - {\overline {I_+({\cal P})}}$, where ${\cal P}$
	is the point $(u,v) = (1/2,1/2)$.
\bibitem{fla-caveat1}
	This statement assumes that the initial slice is outside the
	event horizon.  When considering classically stable
	Reissner-N\"ordstrom-de Sitter spacetimes ($\kappa_c >
	\kappa_i$) in Sec.~\ref{fla-RNdeS}, we
	shall take the rightmost half of our initial data slice
	$\Sigma$ to be a null line outside the event horizon as
	depicted in Fig.~\ref{fla-fig3}; in this
	case the blueshift factor is finite.  If one took the
	rightmost half of the initial slice to be the event horizon
	itself, then the blueshift factor would be infinite, but
	the spacetime would still be (classically) stable because of the
	falloff rate of the initial data on the event horizon.
\bibitem{fla-noteBF}
	Suppose that one uses null coordinates $(u,v)$ such that
	${\vec l} \propto \partial / \partial v$ and ${\vec k} \propto
	\partial/ \partial u$.  Then one might think that the quantity
	$\Psi_{\cal P}({\cal R})$ would depend on the four coordinate values
	$u({\cal P})$, $v({\cal P})$, $u({\cal R})$ and $v({\cal R})$.
	However, $v({\cal P})$ is fixed by the requirement that ${\cal P}$ lie
	on the Cauchy horizon, and $u({\cal R}) = u({\cal P})$ by
	construction.  Thus, $\Psi_{\cal P}({\cal R})$ depends only on two
	independent real parameters.
\bibitem{fla-note6}
	This does not contradict the fact that the integrand is finite
	on the Cauchy horizon itself near ${\cal P}$.  As a function
	of the affine parameter $\lambda$ and the retarded time coordinate
	$v$, the integrand in Eq.~(\ref{fla-figureofmerit}) at large $v$ is
	$l^a \nabla_a R \,\propto \,  f[r(\lambda)] \,
	f^{\prime\prime\prime}[r(\lambda)] \, \exp[\kappa_i
	v]$, with $r(\lambda) \approx r_i + \exp[-\kappa_i v] \lambda$.
	Therefore in the limit $v \to \infty$, $l^a \nabla_a R \to
	f^\prime(r_i) f^{\prime\prime\prime}(r_i) \, \lambda$, which is finite
	at each fixed $\lambda$ but which grows linearly without limit as one
	moves down the Cauchy horizon.  A similar result is found on
	the cosmological horizon.
\bibitem{fla-noteCNG}
	Note that in four dimensional contexts, one needs to
	distinguish between smoothly closed null geodesics, and
	self-intersecting null geodesics in which the direction of the
	tangent to the geodesic	is discontinuous at some point.  In
	the two dimensional context, however, no such distinction is
	necessary, if we assume that the spacetime $(M^\prime,
	g_{ab}^\prime)$ is orientable and time orientable.  This is
	because given a future directed, null,
	tangent vector $k^a$ at a point on a self-intersecting null
	geodesic, when one parallel transports this vector
	around the geodesic, in two dimensions the result must be
	proportional to one of the four null directions at that point.
	The three possibilities other than $k^a$ are excluded by
	orientability and time orientability.
\bibitem{fla-MisnerUnstable}
	The classical instability of Misner space follows from the
	general arguments of Sec.~\ref{fla-blueshiftsec}.  However, there
	is in addition a well-known direct proof that Misner space is
	classically unstable \cite{fla-Yurtsever91}: The general
	solution of the
	wave equation for a massless scalar field in two dimensional Minkowski
	space can be written as $\Phi(u,v) = F(u) + G(v)$ for some functions
	$F$ and $G$, and thus the general solution on Misner space is of this
	form where $F$ and $G$ satisfy
	$$
	F(u) + G(v) = F(e^\xi u) + G(e^{-\xi} v),
	$$
	where $\xi$ is a fixed parameter.  By differentiating this
	equation with respect to $u$ it can
	be shown that the general solution is of the form
	$$
	\Phi(u,v) = A + B \ln (uv) + f[ \ln(-v)] + g[\ln(-u)],
	$$
	where $A$ and $B$ are constants, and $f$ and $g$ are periodic
	functions with period $\xi$.  All of these solutions give rise to a
	diverging stress tensor at $u=0$ or $v=0$ except for the constant
	solution.
\end{references}
\end{document}